\documentclass[%
preprint,
amsmath,amssymb,
aps,
]{revtex4-2}
\usepackage[utf8]{inputenc} 
\usepackage[T1]{fontenc}    
\usepackage{amsfonts}       
\usepackage{booktabs}       
\usepackage{hyperref}       
\usepackage{url}            
\usepackage{microtype}      
\usepackage{nicefrac}       
\usepackage{paralist}       
\usepackage{graphicx}
\usepackage{float}
\usepackage{wrapfig}
\usepackage[dvipsnames]{xcolor}
\usepackage{ulem}
\usepackage{tikz}
\usetikzlibrary{arrows,shapes,angles,quotes}
\usetikzlibrary{decorations.pathreplacing}
\usepackage{fullpage}
\usepackage{rotating}
\usepackage{lmodern}
\usepackage{ulem}
\usepackage{adjustbox}
\usepackage{blindtext}

\usepackage{dcolumn}

\begin{document}

	\title{Structure-Dynamics Relationship in Al-Mg-Si Liquid Alloys}
	\author{Alaa Fahs}
	\affiliation{Univ. Grenoble Alpes, CNRS, Grenoble INP, SIMaP, F-38000 Grenoble, France}
	\affiliation{C-TEC, Parc Economique Centr'alp, 725 rue Aristide Bergès, CS10027, Voreppe 38341 cedex, France }
	
	\author{Philippe Jarry}
	\affiliation{Univ. Grenoble Alpes, CNRS, Grenoble INP, SIMaP, F-38000 Grenoble, France}
	\affiliation{C-TEC, Parc Economique Centr'alp, 725 rue Aristide Bergès, CS10027, Voreppe 38341 cedex, France }
	\author{No\"{e}l Jakse}
	\affiliation{Univ. Grenoble Alpes, CNRS, Grenoble INP, SIMaP, F-38000 Grenoble, France}

	\begin{abstract}
	
	Enhancing properties and performances of aluminium alloys by a control of their solidification
	is pivotal in automotive and aerospace industries. 
	The fundamental role of the structure-diffusion relationship is investigated for Al-Mg-Si liquid alloys taken as a prototype of Al-6xxx. 
	For this purpose, first principles-based molecular dynamics simulations were performed for various Si and Mg content  for Al-rich compositions, including the binary alloy counterparts.
	Results indicate that Mg and/or Si in alloys create a more compact ordering around Al than in pure Al, lowering diffusion. 
	Mg promotes icosahedral short-range order, while Si displays a preference towards cubic local ordering, impacting diffusion based on their respective content. It suggests a
	mechanism whereby an increase in Mg content generally lowers the diffusion of each species, whereas an increase in Si content enhances their diffusion, providing insights for future alloy design.
	
\end{abstract}	
	 
    \maketitle
    \newpage
	
	\section{Introduction}
	Aluminium alloys represent one of the main categories of structural metallic materials
	widely utilized in automotive construction and the aerospace industries \cite{Holmestad2012,Zandbergen1997,Ravi2004,Froseth2003,Jarry2018}. They are
	attractive due to their high strength-to-density ratio, functional extrudability, age hardening
	characteristics, excellent corrosion resistance, as well as surface and welding properties.
	Moreover, the possibility of an almost complete recycling is a strong argument from an
	environmental point of view. Controlling properties and improving quality and performance
	of Al-based alloys represent timely and major industrial challenges \cite{Lassance2006,Mageto2010}.
	
	Improvements can be possible through controlled modification of the targeted microstructure. 
	This requires a good understanding of solidification, a non-equilibrium process governed by the trade-off between solute diffusion in the liquid phase, crystal-liquid interfacial energy and anisotropy \cite{Dantzig2016}.
	Unfortunately, experimental data for self-diffusion in liquid metals are scarce and most of them related to monatomic liquid metal, mainly due to a lack of specific radio-isotopes \cite{Iida1988}. 
	Alternatively, inserting the experimental viscosity data, widely available in the literature, into phenomenological laws such as the Stokes-Einstein (SE) equation \cite{Sutherland1905,Einstein1905}, could indirectly give a rough estimation of solute diffusivity. 
	However, this approximated method happens to fail in most of the cases for multi-component liquid alloys. It might be due to a possible different local ordering around each species \cite{Jakse2016,Pasturel2016}, which falls out of the assumption of the SE relation, and does not extend to undercooling conditions where solidification occurs.
	
	Similarities of densities and coordination numbers of the liquid and crystal for instance led to recognise that the local atomic ordering, so-called the short-range order (SRO), should be somewhat incompatible with the long range crystalline structure \cite{Turnbull1952}. 
	Frank \cite{Frank1952} was the first to show that the icosahedral ordering is locally more stable. This might engender complex situations in the undercooled melt in terms of the variety of polymorphs \cite{Ronceray2011}, competing short range orders \cite{Jakse2003,Becker2020,Pedersen2021}, or an interplay between chemical and fivefold symmetry (FFS) orderings in liquid alloys \cite{Jakse2008,Pasturel2017,Tang2018} prior to nucleation. More precisely, the proportion of atom pairs belonging to Icosahedral Short-Range Order (ISRO) strongly depends upon the nature of the solutes. In particular, in the Al$_{93}$Cr$_7$ liquid alloy, \textit{ab initio} molecular dynamics (AIMD) calculations show the existence of an onset of Icosahedral Medium-Range Order (IMRO), induced by the interplay between Chemical Short-Range Order (CSRO) and ISRO. This leads to dynamical heterogeneities characterized by a strong heterogeneity in the propension to diffuse \cite{Pasturel2017,Pasturel2017*}. Another AIMD study of the liquid Al$_{1-x}$Ni$_x$ alloy pointed out the existence of a non-linear evolution of dynamic properties due to non-linear composition dependence of CSRO \cite{Jakse2015}. More generally, a deeper understanding of the structure-dynamic relationship in liquid metals and alloys is  needed for designing alloys with desired properties.
	
	In this context, the possibility of calculating dynamic properties together with the underlying structure in the liquid and undercooled states on the same footing can be provided from \textit{ab initio} molecular dynamics \cite{Kurtuldu2013,Kurtuldu2014,Kurtuldu2015,Rappaz2020,Jakse2016,Pasturel2016,Jakse2014,Jakse2017,Bouhadja2019,Jakse2008,Pasturel2010,Pasturel2017,Pasturel2017*,Jakse2015,Jakse2013}. For most metallic systems, AIMD is able to accurately account for interatomic interactions on the basis of the Density Functional Theory (DFT) \cite{Hafner2008}. 
	Despite the heaviness of the underlying quantum computation to get the electronic structure, AIMD simulations currently allow us to build phase space trajectories of system sizes of the order of 10$^2$ atoms over several hundreds of picoseconds \cite{Pasturel2017}, relevant for the determination of diffusion coefficients for all species. 
	Another strategy consists in developing Machine Learning (ML) interatomic potentials trained on an appropriate set of AIMD configurations, allowing to perform up to million-atoms molecular dynamics (MD) simulations with an accuracy close to \textit{ab initio} \cite{Behler2021}.
	It should be noted that ML potentials for aluminium and its alloys in the liquid and solid states were proposed recently \cite{Smith2021,Jakse2023,Jain2021} and used to study early stage of solidification \cite{Jakse2023}.\\
	\indent The present work aims at investigating the structure-dynamic relationships in liquid and undercooled Al-Mg-Si ternary alloys, which are prototypical to the Al-based 6xxx used in the automobile industry. 
	For this purpose, \textit{ab initio} molecular dynamics are performed for various Al-rich compositions, namely, Al\textsubscript{80}Mg\textsubscript{10}Si\textsubscript{10}, Al\textsubscript{70}Mg\textsubscript{20}Si\textsubscript{10}, and Al\textsubscript{70}Mg\textsubscript{10}Si\textsubscript{20}. In particular, the addition of Mg and Si in liquid aluminium is further quantified by performing also AIMD simulations for Al\textsubscript{90}Mg\textsubscript{10} and the Al\textsubscript{90}Si\textsubscript{10} binary counterparts to disentangle effects of Mg and Si in the ternary liquid alloys. Our results show significant structural heterogeneities with Mg or Si addition in aluminium either in binary or ternary alloys. In all cases, an enhancement of the FFS around Mg atoms is observed, while a decrease is seen around Si, leading to a decoupling of diffusion. Impact of these findings on the early stages of solidification in the alloys is further discussed. \\
	\indent The layout of the paper is the following: Sec. II is devoted to the computational aspects of
	the work, Sec. III shows results of structural and dynamic properties along with a discussion
	of the interplay between the local structural properties and the dynamics, and finally, Sec IV
	draws the conclusions.
	
	\section{Computational background}
	
	\subsection{Ab initio molecular dynamics}
	
	
	The structure and dynamic properties of binary Al\textsubscript{90}Mg\textsubscript{10} and Al\textsubscript{90}Si\textsubscript{10}, as well as ternary Al\textsubscript{80}Mg\textsubscript{10}Si\textsubscript{10}, Al\textsubscript{70}Mg\textsubscript{20}Si\textsubscript{10}, and Al\textsubscript{70}Mg\textsubscript{10}Si\textsubscript{20} liquid alloys, were determined by means of \textit{ab initio} molecular dynamics simulations within the density functional theory, using the Vienna \textit{Ab initio} Simulation Package (VASP) \cite{Kresse1993}. Projected augmented-wave method was used to describe the electron-ion interaction \cite{Kresse1999}, with a plane wave cutoff of $250$\,eV and only the $\Gamma$-point to sample the Brillouin zone. Exchange and correlation effects were taken into account through the Local Density Approximation (LDA)\cite{Ceperley1981}. As a matter of fact, it was shown in preceding works that the LDA gives a satisfactory representation of pure Al and Al-rich alloys in the liquid and undercooled states\cite{Jakse2013,Pasturel2017*,Bryk2018,Demmel2021}.
	
	The phase space trajectories were constructed through dynamical simulations by numerical integration of Newton's equation of motion using the standard Verlet algorithm \cite{Verlet1967} in the velocity form with a time step of $1.5$\,fs. 
	All simulations were done in a cubic box of $N = 256$ atoms with periodic boundary conditions in the three directions of space. The number of each species in all the simulations can be found in Table I of the Supporting Information file. 
	For each simulation, the volume $V$ was adjusted to reach pressure with an uncertainty less than $0.5$ GPa, which is of the order of pressure fluctuations (see Fig. S1 in the SI file). 
	In order to control the temperature $T$, the simulations were performed in the canonical ensemble (constant $N$, $V$, and $T$) with a Nos\'e thermostat \cite{Nose1984}. 
	Thermalization was carried out for at least $20$\,ps. Simulations were continued during $60$\,ps to extract the structural and dynamic properties. The temperature evolution in the undercooled states was obtained by quenching the system step-wise down to $600$\,K with a temperature step of $100$\,K from an equilibrated state at the previous temperature, resulting in an average quenching rate of $10^{13}$\,K/s.
	
	From equilibrated trajectories at each temperature, a regular sampling of $10$ configurations was performed to obtain Inherent Structure (IS) configuration \cite{Stillinger1982}, from which a structural analysis is extracted by means of the the Common-Neighbor Analysis (CNA)  \cite{Faken1994} using Python modules of the Open VIsualization TOol (OVITO) \cite{Stukowski2010}.
	To this end, static relaxation simulations were performed for each of these configurations, performing an energy
	minimization procedure with the conjugate gradient method.
	This procedure is important to uncouple the vibrational motion from the underlying structural properties by bringing atoms to a local minimum in the potential-energy surface and reduce the noise in the CNA index calculations.

	\subsection{Structural and dynamic properties}

	Structural properties were first investigated through the calculation of the total structure factor using the Debye equation for the purpose of comparing the simulation results to existing experimental measurements:
	
	\begin{equation}
		S(q)=\cfrac{1}{N} \left< \left| \sum_{i} e^{\rm{i} \textit{\textbf{q}}\cdot\textit{\textbf{r\textsubscript{i}}} } \right|^2 \right> ~,
		\label{eq.1}
	\end{equation}
	where $\textit{\textbf{q}}$ is the wave vector, $\textit{\textbf{r\textsubscript{i}}}$ the atomic position, and the angle brackets denote averaging over configurations sampled along the phase-space trajectory. The partial structural factors were calculated in the same manner within the Faber-Ziman formalism \cite{Faber1965}.
	
	The partial pair-correlation functions $g_{ij}(r)$ were also also considered to give a view of the local structural properties in the real space. These functions are written as:
	
	\begin{equation}
		g_{ij}(r) = \cfrac{V}{N_{i} 4 \pi r^2} \lim_{\Delta r \rightarrow 0} \cfrac{n_{ij}(r+\Delta r)}{\Delta r} ~,
		\label{eq.2}
	\end{equation}
	where $i$ and $j$ refer to two different types of atoms, $N_{i}$ is the number of atoms of type $i$, $n_{ij}(r)$ is the mean number of atoms of type $j$ around an atom of type $i$ in a spherical shell of radius $r$ and thickness $\Delta r$ centered on an atom $i$ taken as the origin. The corresponding partial coordination number ($N_c$) or the number of first nearest neighbors of a reference atom was calculated using the following equation:
	
	\begin{equation}
		N_{ij} = \rho_j \int_{0}^{r_{min}} 4 \pi r^2 g_{ij}(r) {\rm d}r~,
		\label{eq.3}
	\end{equation} 
	where $\rho_j = N_{j}/V$ is the partial number density, and $r_{min}$ corresponds to the distance at the first minimum of the  corresponding partial pair-correlation function $g_{ij}(r)$.
	
	The dynamics of an atom of type $i$ can be tracked by means of the mean-square-displacement as function of time $t$, which is written as follows:
	
	\begin{equation}
		R^2_i(t)= \cfrac{1}{N_i} \sum_{k=1}^{N_i}\left< \left[ \textbf{r}_k(t+t_0)-\textbf{r}_k(t_0)\right]^2 \right>_{t_0}~,
		\label{eq.4}
	\end{equation} 
	where the angular brackets represent an averaging over time origins $t_0$.  Then, partial diffusion coefficients $D_i$ could be deduced from $R^2_i(t)$ as follows:
	
	\begin{equation}
		D_i= \lim\limits_{t \rightarrow +\infty} \cfrac{R^2_i(t)}{6t}~.
		\label{eq.5}
	\end{equation}
	
	\subsection{Comparison with experiments}
	
	\begin{figure*}[t]
		\begin{center}
			\includegraphics[angle=0 ,width=1\textwidth]{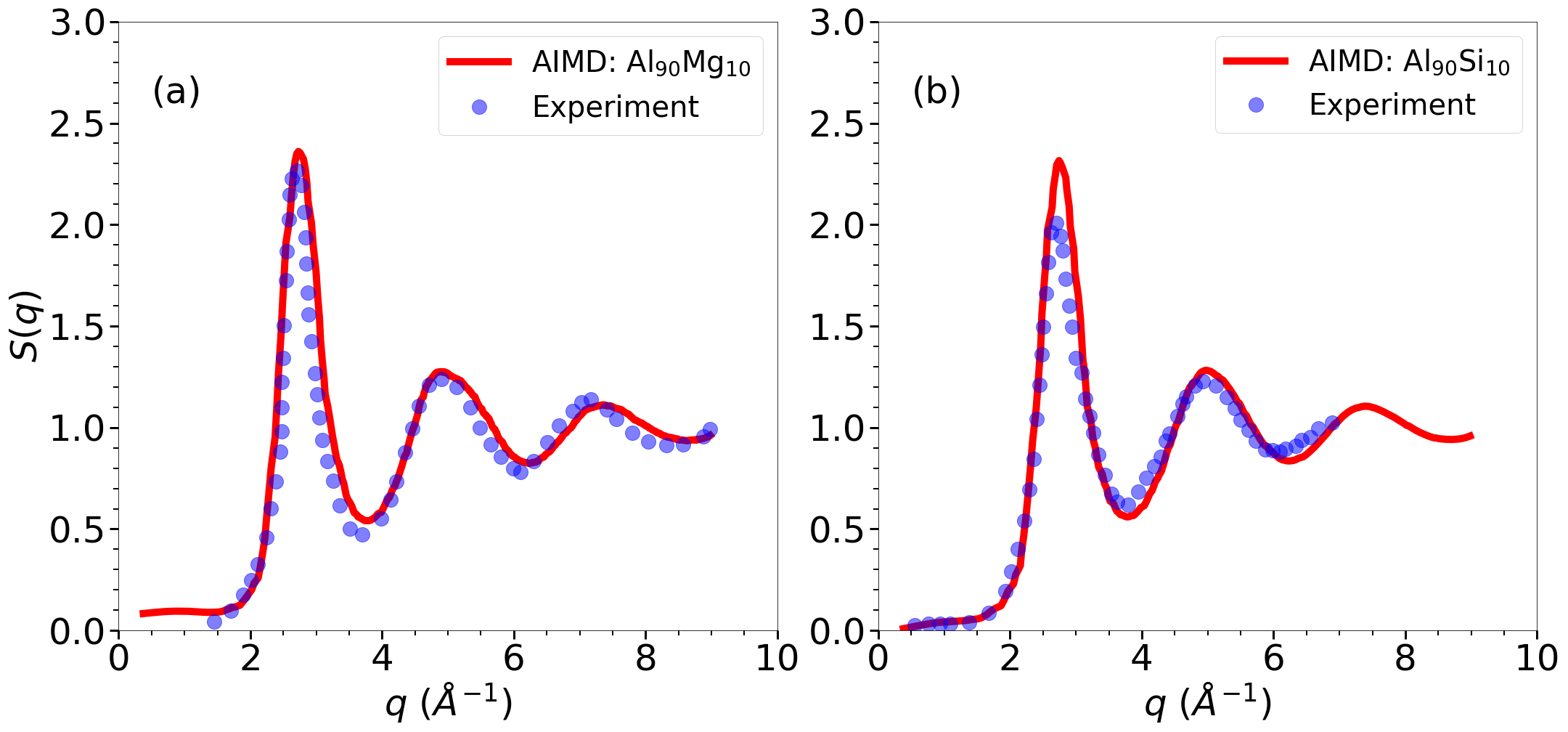}

		\end{center}
		\caption{AIMD calculations of the total X-ray structure factors in liquid Al$_{90}$Mg$_{10}$ (a) and Al$_{90}$Si$_{10}$ (b) (solid lines), at temperature $T = 1000$ K and $T = 900$ K, respectively. Experimental data are shown with the symbols \cite{Waseda1973,Shtablavyi2008} for comparison.}
		\label{fig1}   
	\end{figure*}
	
	First of all, the reliability of our \textit{ab initio} calculations is assessed by comparing calculated total structure factors $S(q)$ with existing experimental data \cite{Waseda1973,Shtablavyi2008}. Figs. 1 (a) and (b) show AIMD results for Al$_{90}$Mg$_{10}$ and Al$_{90}$Si$_{10}$ liquid alloys at temperatures $T = 1000$ K and $T = 900$ K, respectively. A good agreement is obtained with X-ray diffraction data \cite{Waseda1973,Shtablavyi2008} regarding the position and amplitude of the first peak and subsequent oscillations. Moreover, Al-Mg and Mg-Mg partial structure factors in liquid Al$_{90}$Mg$_{10}$ are compared to experimental data of Waseda \cite{Waseda1973} in Fig. S2 in the SI file within the Faber-Ziman formalism \cite{Faber1965}. The Mg-Mg partial structure factor is well reproduced. This is all the more true for Al-Mg partial, especially regarding the peak positions, even if their amplitude is a bit underestimated. Such a result is particularly remarkable given the experimental difficulties to extract partials. To the best of our knowledge, no experimental data were reported in the literature for the ternary Al-Mg-Si liquid alloys.
	
	\subsection{Comparison with machine learning potentials}
	
	Machine learning interatomic potentials were developed recently for the Al-Mg-Si ternary system \cite{Jain2021} using a High Dimensional Neural Network Potential (HDNNP) \cite{Behler2007}. Using the version named NN-19, extracted from the Materials Cloud \cite{MaterialsCloud}, local structural and diffusion properties were carried out by molecular dynamics (MD) using LAMMPS in the same conditions as our AIMD calculations. The use of HDNNP was possible by means of the library-based implementation n2p2 by Singraber \textit{et al.} \cite{Singraber2019}.
	
\begin{figure*}[]
		\begin{center}
			\includegraphics[angle=0 ,width=1\textwidth]{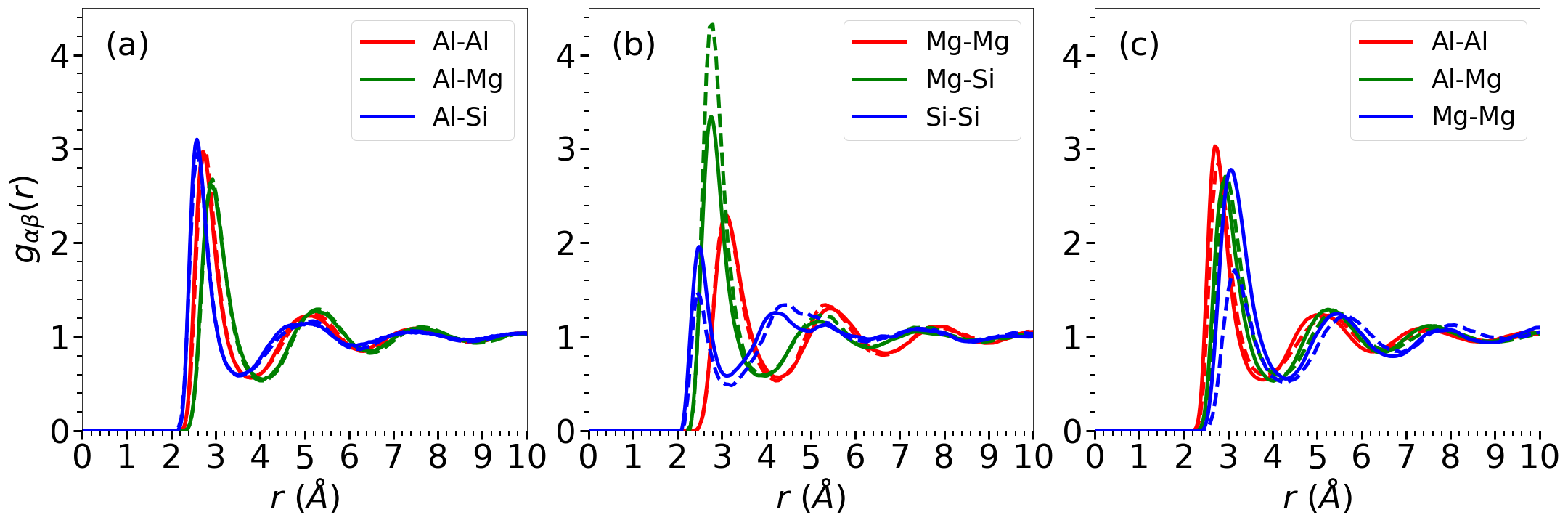} 
			
		\end{center}
		\caption{Partial pair-correlation functions of Al\textsubscript{80}Mg\textsubscript{10}Si\textsubscript{10} (a) and (b) as well as Al\textsubscript{90}Mg\textsubscript{10} (c) at
			$T = 1000$ K. HDNNP results (dashed lines) are compared to AIMD data (solid lines).}
		\label{fig:gternaryNN}   
	\end{figure*}

	Partial pair-correlation functions of Al\textsubscript{80}Mg\textsubscript{10}Si\textsubscript{10} liquid alloy from HDNNP at $T = 1000$ K are plotted in Figs. \ref{fig:gternaryNN} (a) and (b). As a general trends, AIMD curves are well reproduced by the HDNNP, especially for Al-Al, Al-Mg and Al-Si pairs. Only the first peak of the Mg-Si partial shows an overestimated amplitude, and thus a higher Mg-Si affinity. More importantly, diffusion coefficients shown in Fig. S3 in the SI file, are in good agreement with AIMD in the liquid and undercooled states. 
	This demonstrates the reliability of this potential and opens the way to larger scale simulations for early stage of solidification, with close to \textit{ab initio} accuracy in the range of compositions studied here. 
	The  Al\textsubscript{90}Mg\textsubscript{10} and Al\textsubscript{90}Si\textsubscript{10} binary counterparts of the Al-Mg-Si ternary system were also investigated with the HDNNP. 
	Fig. \ref{fig:gternaryNN} (c) comparison with AIMD for Al\textsubscript{90}Mg\textsubscript{10}, where a good agreement is seen. However, the self-diffusion coefficients reported in Fig. S4 show increasing departure with temperature, yet keeping the correct hierarchy $D_{\rm Mg}$ < $D_{\rm Al}$. 
	For Al\textsubscript{90}Si\textsubscript{10} it was not possible run stable MD simulations in the liquid state. 
	The HDNNP for the ternary Al-Mg-Si ternary system is seemingly not completely transferable to binary counterparts. Training set should include the corresponding binary alloys, and the monoatomic metals if it were to be used for studying solidification phenomena.
	
	\begin{figure*}[]
		\begin{center}
			\includegraphics[angle=0 ,width=1\textwidth]{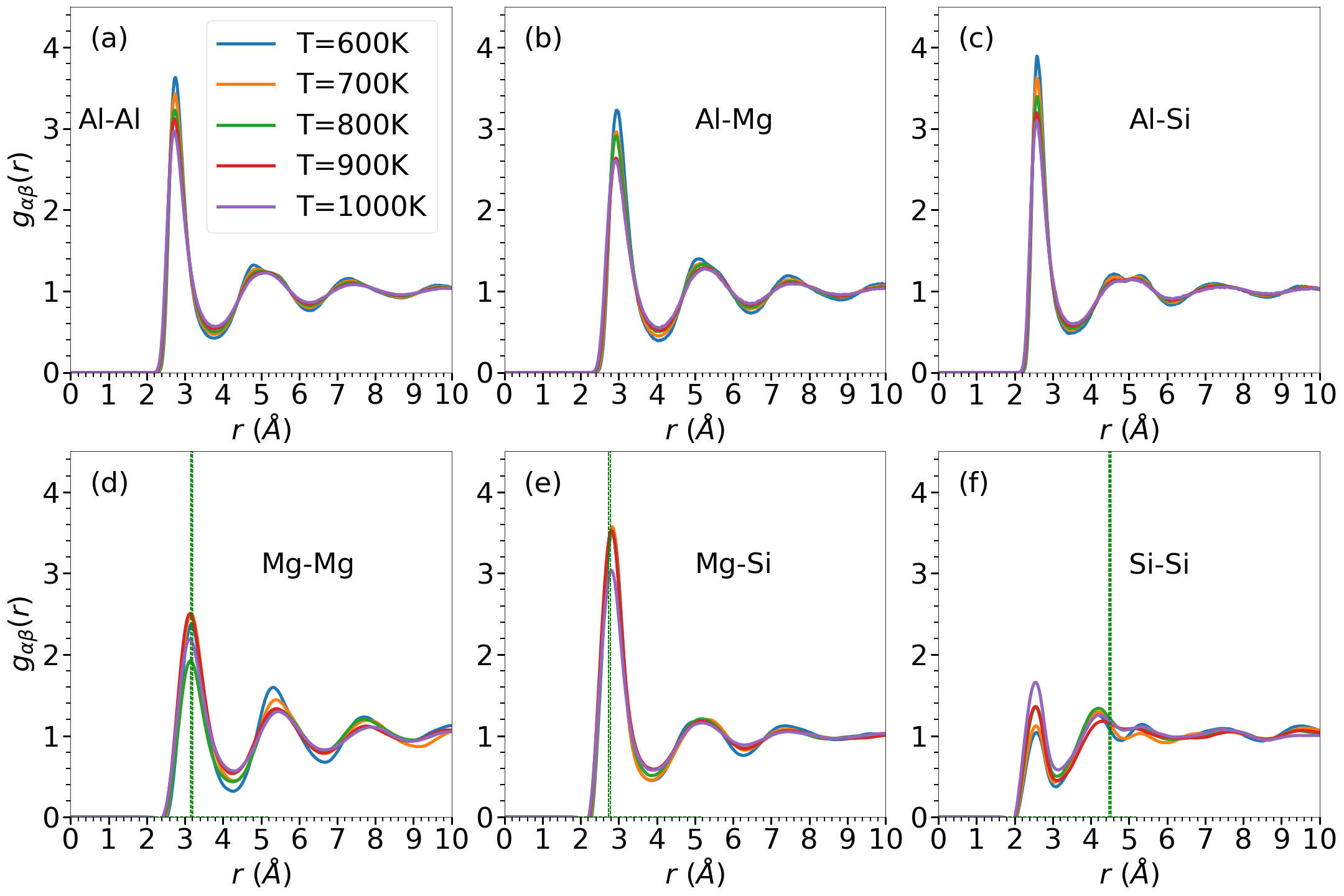}	
			
		\end{center}
		\caption{Partial pair-correlation functions of Al$_{80}$Mg$_{10}$Si$_{10}$ liquid alloy for all studied temperatures by AIMD simulations. The green dashed vertical line indicates the first atomic distances of pairs in the Mg$_2$Si cubic crystal (The Materials Project \cite{Materialsproject}: reference mp-1367).}
		\label{fig:gternary}   
	\end{figure*}
	
	\section{Results}
	\subsection{Structural properties}

	The local structural properties are first considered through the partial pair correlation functions calculated from the AIMD simulations using Eq. \ref{eq.2}. They are shown in Fig. \ref{fig:gternary}
	(a)-(f) for the Al$_{80}$Mg$_{10}$Si$_{10}$ ternary liquid alloy. The first peak position of each partial reflects the first neighbor average distances, and thus the bonding length between the corresponding type of pairs. Going from the liquid state close to the liquidus at $T = 1000$ K to the undercooled states at $T = 600$ K, the first peak does not change significantly in its position whatever the partial. Nevertheless, its height increases visibly, showing that the short-range order becomes more and more pronounced and well defined. This is all the more true for the second oscillation whose amplitude increases and eventually splits in two sub-peaks, a feature that is often attributed to the reinforcement of the FFS \cite{Schenk2002,Pasturel2010,Jakse2013,Jakse2015,Royall2015,Pasturel2017}. Fig. \ref{fig:gternary}(f) shows that the amplitude of the $g_{\rm SiSi}(r)$ remains very small and close to unity, indicating that Si atoms are randomly dispersed in the Al host liquid. 
	Table II in the SI file
	gathers the first peak position value of all the partials. It is worth mentioning first that addition
	of a small amount of Mg and Si in pure Al leaves the first nearest distance unchanged with
	$r_{\rm AlAl}$ = 2.7~\AA, close to pure liquid Al \cite{Jakse2013,Mauro2011,Alemany2004}. It also shows that $r_{\rm AlMg}$ is intermediate
	between $r_{\rm AlAl}$ and $r_{\rm MgMg}$. The height of the first peak of $g_{\rm AlMg}(r)$ is also intermediate. The
	same holds for $r_{\rm AlSi}$, but in that case, the first peak of $g_{\rm AlSi}(r)$ is higher than $g_{\rm AlAl}(r)$ and
	$g_{\rm SiSi}(r)$, showing a slightly higher affinity between Al and Si.

	\begin{figure}[t]
		\begin{center}

			\includegraphics[angle=0 ,width=0.6\textwidth]{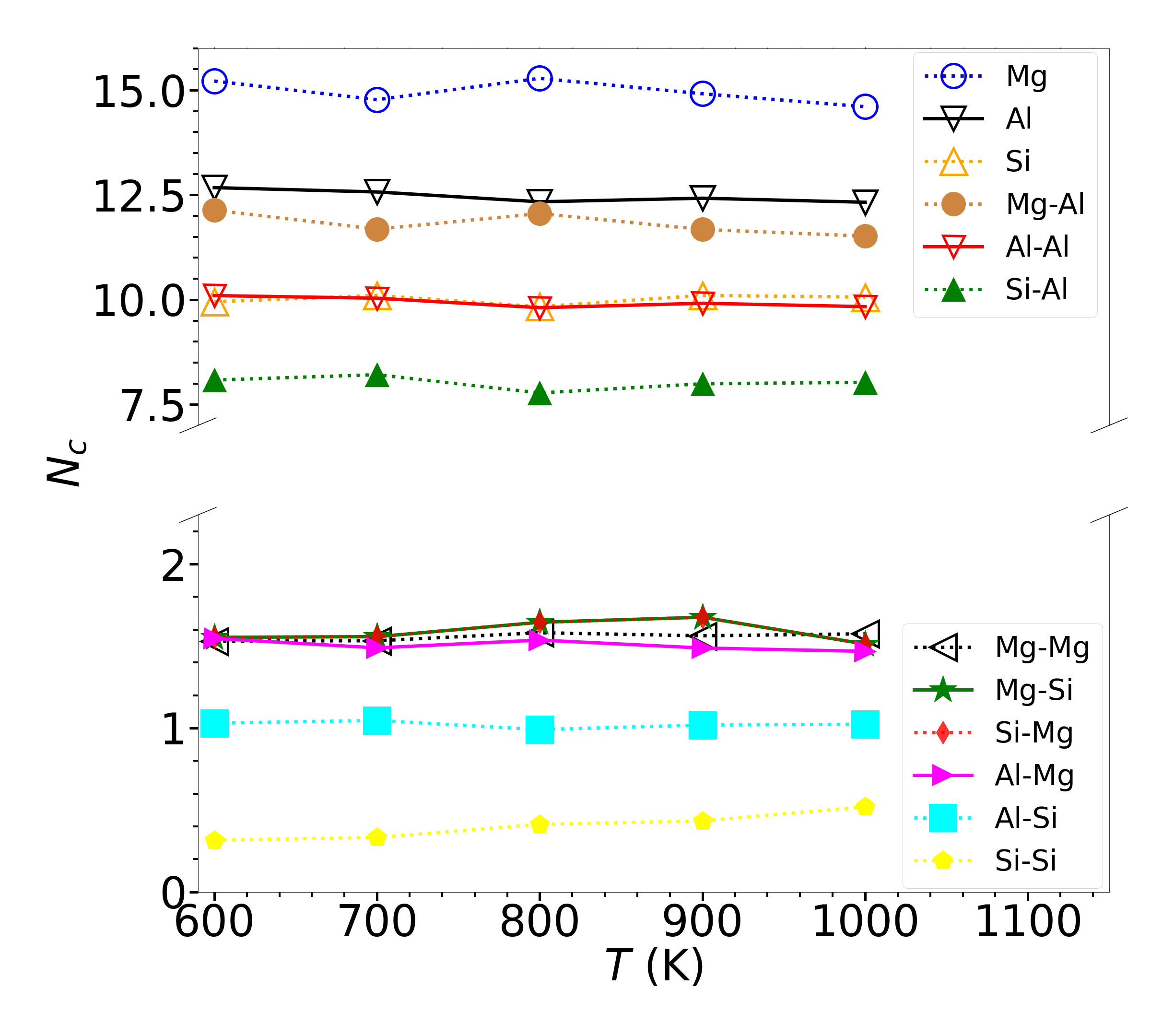}

		\end{center}
		\caption{Total and partial coordination numbers as function of temperature in the ternary Al$_{80}$Mg$_{10}$Si$_{10}$ liquid alloy. Uncertainties in the values are of the order of $0.2$.}

		\label{fig3}   
	\end{figure}

	In order to investigate a potential ternary effect, AIMD simulations were performed for two
	binary alloys, namely Al$_{90}$Mg$_{10}$ and Al$_{90}$Si$_{10}$ in the same range of temperatures. The results
	for the partial pair-correlation functions are shown in Figs. S7 and S8, and the corresponding
	bond lengths are included in Table II in the SI file. The same trends on the partials are seen,
	and only little change can be observed with respect to the ternary alloy. Nevertheless,
	addition of Si in Al-Mg binary alloys leads to the occurrence of a significant Mg-Si bonding
	in the ternary alloy as can be seen in Fig. \ref{fig:gternary}(e). A similarity can be observed between the calculated
	Mg-Mg or Mg-Si distances and those reported for the Mg$_2$Si crystal structure, whose first neighbor distances are indicated by the vertical dashed lines in Fig. \ref{fig:gternary} (d)-(f). The bond-angle
	distribution in Fig. S9 of the SI file also suggests short-range ordering close to this phase.
	
	The resulting partial and total coordination numbers and their temperature evolution are represented in Fig. \ref{fig3} for Al$_{80}$Mg$_{10}$Si$_{10}$. 
	Only a modest increase of the total coordination numbers around Al, Mg and Si is observed with decreasing temperature between $T=1000$\,K and $T=600$\,K.   
	The total coordination number of Al is: $N_c^{\rm Al}$ = 12.4 $\pm~0.2$, which is an intermediate value between $N_c^{\rm Mg}=$ 15.0 $\pm~0.2$ and $N_c^{\rm Si}=$ 10 $\pm~0.2$. Similar values for the total coordination numbers were found for the two binary liquid alloys (see Fig. S10 in the SI file). 
	For all liquid alloys (binary and ternary), the total coordination numbers show the following trend: $N_c^{\rm Si}< N_c^{\rm Al}<N_c^{\rm Mg}$. Moreover, addition of 10\% Mg and/or 10\% Si leads to a significant increase of coordination number around Al with respect to pure Al \cite{Jakse2013,Alemany2004}. This feature implies (i) an increase of the compacity since the coordination sphere remains the same as mentioned above, and (ii) the possible appearance of a specific chemical short-range order. 
	The following inequality, $N_c^{\rm Si}<N_c^{\rm Al}<N_c^{\rm Mg}$, could indicate that, in general, the local structure around each species is different, characteristic of a structural heterogeneity. It is worth noting that Mg-Mg and Mg-Si partial coordination numbers are larger than 1, and for Al-Si it is equal to 1, whereas Si-Si coordination is almost negligible. 
	This confirms that, in contrast with Si-Si bonding, Mg-Mg pairing is favored in both Al$_{90}$Mg$_{10}$ and Al$_{80}$Mg$_{10}$Si$_{10}$, and that Si atoms exhibit preferential coordination with Mg, as expected by the analysis of the corresponding partial pair-correlation function. 
	This is characteristic of the appearance of a CSRO confirmed by means of the Warren-Cowley parameter \cite{Cowley1950}, $\alpha_i=1-N_c^{i-\rm Al}/x_{\rm Al}(N_c^{i-i}+N_c^{i-\rm Al})$, giving $\alpha_{\rm Si}=-0.06$ in Al$_{90}$Si$_{10}$ and $\alpha_{\rm Mg}=0.04$ in Al$_{90}$Mg$_{10}$.

	\begin{table*}[th!]
		
		\caption{Common Neighbor Analysis for five Al-based liquid alloys calculated at $T=1000$ K and $T=600$ K. Results for pure liquid Al, reported in Ref. \cite{Jakse2013}, are also shown for comparison. The typical error bar for the calculated pair abundances are of the order of $0.01$.}
		\begin{ruledtabular}
			\begin{tabular}{ccccccc}
				
				
				CNA (\%) 
				& Al \cite{Jakse2013}
				& Al$_{90}$Mg$_{10}$ 
				& Al$_{90}$Si$_{10}$ 
				& Al$_{80}$Mg$_{10}$Si$_{10}$ 
				& Al$_{70}$Mg$_{20}$Si$_{10}$ 
				& Al$_{70}$Mg$_{10}$Si$_{20}$\\
				
				\hline
				
				$T=1000$K
				&  \hspace{0.05cm}  
				& Al \hspace{0.4cm} Mg 
				& Al \hspace{0.4cm} Si 
				& Al \hspace{0.4cm} Mg \hspace{0.4cm} Si 
				& Al \hspace{0.4cm} Mg \hspace{0.4cm} Si 
				& Al \hspace{0.4cm} Mg \hspace{0.4cm} Si\\
				\hline
				\vspace{0.05cm}
				$[5~5~5] + [5~4~4]$ & 26.9 
				& 38.1 \hspace{0.2cm} 45.1 
				& 29.3 \hspace{0.2cm} 12.6 
				& 32.9 \hspace{0.2cm} 39.3 \hspace{0.2cm} 15.6 
				& 36.3 \hspace{0.2cm} 40.5 \hspace{0.2cm} 20.9 
				& 31.0 \hspace{0.2cm} 37.3 \hspace{0.2cm} 13.9\\
				\vspace{0.05cm}  
				
				$[6~~~6~~~6]$        & 1.40 
				& 2.83 \hspace{0.2cm} 5.39 
				& 2.70 \hspace{0.2cm} 0.53 
				& 2.33 \hspace{0.2cm} 5.57 \hspace{0.2cm} 0.49  
				& 2.78 \hspace{0.2cm} 7.12 \hspace{0.2cm} 0.87  
				& 2.51 \hspace{0.2cm} 5.53 \hspace{0.2cm} 0.44\\
				\vspace{0.05cm}  
				
				$[4~~~4~~~4]$        & 1.40  
				& 2.24 \hspace{0.2cm} 1.03 
				& 2.04 \hspace{0.2cm} 2.05 
				& 1.82 \hspace{0.2cm} 1.24 \hspace{0.2cm} 2.38  
				& 2.68 \hspace{0.2cm} 2.14 \hspace{0.2cm} 4.01 
				& 2.01 \hspace{0.2cm} 1.16 \hspace{0.2cm} 2.06\\
				\vspace{0.05cm} 
				
				$[4~~~3~~~3]$        & 20.0 
				& 25.0 \hspace{0.2cm} 21.1 
				& 23.6 \hspace{0.2cm} 25.9 
				& 24.5 \hspace{0.2cm} 21.6 \hspace{0.2cm} 27.0 
				& 26.2 \hspace{0.2cm} 21.4 \hspace{0.2cm} 32.0  
				& 23.4 \hspace{0.2cm} 20.7 \hspace{0.2cm} 25.5\\
				\vspace{0.05cm} 
				
				$[4~~~2~~~2]$        & 13.9 
				& 11.2 \hspace{0.2cm} 9.58 
				& 13.3 \hspace{0.2cm} 13.6 
				& 13.0 \hspace{0.2cm} 11.1 \hspace{0.2cm} 12.5  
				& 10.0 \hspace{0.2cm} 9.30 \hspace{0.2cm} 9.73  
				& 12.3 \hspace{0.2cm} 10.5 \hspace{0.2cm} 13.0\\
				\vspace{0.05cm} 
				
				$[4~~~2~~~1]$        & 8.00 
				& 6.34 \hspace{0.2cm} 5.71 
				& 6.05 \hspace{0.2cm} 5.74 
				& 5.34 \hspace{0.2cm} 5.18 \hspace{0.2cm} 4.83  
				& 5.24 \hspace{0.2cm} 4.65 \hspace{0.2cm} 4.54  
				& 5.08 \hspace{0.2cm} 5.41 \hspace{0.2cm} 4.89\\
				\vspace{0.05cm} 
				
				[3~~~2~~~2]          & 4.10  
				& 3.34 \hspace{0.2cm} 2.52 
				& 4.36 \hspace{0.2cm} 6.79 
				& 3.90 \hspace{0.2cm} 3.29 \hspace{0.2cm} 7.33 
				& 3.96 \hspace{0.2cm} 3.05 \hspace{0.2cm} 7.01 
				& 4.35 \hspace{0.2cm} 3.39 \hspace{0.2cm} 7.13\\
				\vspace{0.05cm} 
				
				[3~~~1~~~1]          & 10.6  
				& 6.84 \hspace{0.2cm} 4.86 
				& 10.3 \hspace{0.2cm} 19.2 
				& 9.48 \hspace{0.2cm} 5.79 \hspace{0.2cm} 18.3  
				& 7.80 \hspace{0.2cm} 6.03 \hspace{0.2cm} 13.7 
				& 10.4 \hspace{0.2cm} 7.43 \hspace{0.2cm} 18.6\\
				\vspace{0.05cm}

				[3~~~0~~~0]          & -  
				& 1.44 \hspace{0.2cm} 1.10 
				& 3.00 \hspace{0.2cm} 5.66 
				& 2.69 \hspace{0.2cm} 1.45 \hspace{0.2cm} 5.14  
				& 1.45 \hspace{0.2cm} 0.99 \hspace{0.2cm} 3.53 
				& 3.12 \hspace{0.2cm} 1.43 \hspace{0.2cm} 5.78\\
				\vspace{0.05cm} 
				
				[2~~~0~~~0]          & -  
				& 0.66 \hspace{0.2cm} 0.62 
				& 1.52 \hspace{0.2cm} 4.50 
				& 1.23 \hspace{0.2cm} 1.13 \hspace{0.2cm} 3.70  
				& 0.86 \hspace{0.2cm} 0.56 \hspace{0.2cm} 2.22 
				& 1.74 \hspace{0.2cm} 1.01 \hspace{0.2cm} 5.41\\

				\hline
				$T=600$K
				&  \hspace{0.05cm}  
				& Al \hspace{0.4cm} Mg 
				& Al \hspace{0.4cm} Si 
				& Al \hspace{0.4cm} Mg \hspace{0.4cm} Si  
				& Al \hspace{0.4cm} Mg \hspace{0.4cm} Si  
				& Al \hspace{0.4cm} Mg \hspace{0.4cm} Si\\
				\hline
				\vspace{0.05cm} 
				
				$[5~5~5] + [5~4~4]$ & - 
				& 48.8 \hspace{0.2cm} 56.9 
				& 37.3 \hspace{0.2cm} 22.1 
				& 46.1 \hspace{0.2cm} 52.5 \hspace{0.2cm} 24.2 
				& 49.5 \hspace{0.2cm} 51.3 \hspace{0.2cm} 32.2  
				& 39.2 \hspace{0.2cm} 44.4 \hspace{0.2cm} 23.0\\
				\vspace{0.05cm} 
				
				$[6~~~6~~~6]$        & - 
				& 3.23 \hspace{0.2cm} 9.79 
				& 2.39 \hspace{0.2cm} 0.82 
				& 3.67 \hspace{0.2cm} 8.85 \hspace{0.2cm} 0.65  
				& 4.72 \hspace{0.2cm} 10.0 \hspace{0.2cm} 1.21  
				& 4.14 \hspace{0.2cm} 8.35 \hspace{0.2cm} 0.76\\
				\vspace{0.05cm} 
				
				$[4~~~4~~~4]$        & -  
				& 1.85 \hspace{0.2cm} 1.84 
				& 1.70 \hspace{0.2cm} 0.32 
				& 2.14 \hspace{0.2cm} 1.71 \hspace{0.2cm} 2.85  
				& 3.54 \hspace{0.2cm} 2.62 \hspace{0.2cm} 5.96 
				& 2.65 \hspace{0.2cm} 1.78 \hspace{0.2cm} 3.63\\
				\vspace{0.05cm} 
				
				$[4~~~3~~~3]$        & - 
				& 21.8 \hspace{0.2cm} 16.3 
				& 22.8 \hspace{0.2cm} 23.0 
				& 22.4 \hspace{0.2cm} 16.2 \hspace{0.2cm} 31.4 
				& 22.1 \hspace{0.2cm} 18.1 \hspace{0.2cm} 32.8  
				& 23.4 \hspace{0.2cm} 18.5 \hspace{0.2cm} 31.2\\
				\vspace{0.05cm} 
				
				$[4~~~2~~~2]$        & - 
				& 9.74 \hspace{0.2cm} 6.33 
				& 13.0 \hspace{0.2cm} 13.6 
				& 9.96 \hspace{0.2cm} 7.94 \hspace{0.2cm} 12.7  
				& 7.97 \hspace{0.2cm} 6.51 \hspace{0.2cm} 8.46  
				& 10.5 \hspace{0.2cm} 9.59 \hspace{0.2cm} 11.4\\
				\vspace{0.05cm} 
				
				$[4~~~2~~~1]$        & - 
				& 5.96 \hspace{0.2cm} 3.49 
				& 7.07 \hspace{0.2cm} 6.38 
				& 4.56 \hspace{0.2cm} 3.89 \hspace{0.2cm} 5.00   
				& 4.15 \hspace{0.2cm} 3.00 \hspace{0.2cm} 4.20  
				& 4.95 \hspace{0.2cm} 4.25 \hspace{0.2cm} 5.10\\  
				\vspace{0.05cm} 
				
				[3~~~2~~~2]          & -  
				& 2.14 \hspace{0.2cm} 1.02 
				& 3.04 \hspace{0.2cm} 6.53 
				& 2.39 \hspace{0.2cm} 1.44 \hspace{0.2cm} 4.85   
				& 2.13 \hspace{0.2cm} 1.86 \hspace{0.2cm} 4.46   
				& 3.30 \hspace{0.2cm} 1.96 \hspace{0.2cm} 6.12\\ 
				\vspace{0.05cm}

				[3~~~1~~~1]          & -  
				& 4.06 \hspace{0.2cm} 1.41 
				& 7.24 \hspace{0.2cm} 16.6 
				& 4.90 \hspace{0.2cm} 3.15 \hspace{0.2cm} 12.2  
				& 2.95 \hspace{0.2cm} 2.16 \hspace{0.2cm} 7.56  
				& 6.21 \hspace{0.2cm} 4.18 \hspace{0.2cm} 12.2\\
				\vspace{0.05cm} 
				
				[3~~~0~~~0]          & -  
				& 0.78 \hspace{0.2cm} 0.56 
				& 1.54 \hspace{0.2cm} 3.49 
				& 1.21 \hspace{0.2cm} 0.58 \hspace{0.2cm} 2.97  
				& 0.49 \hspace{0.2cm} 0.43 \hspace{0.2cm} 1.06 
				& 1.50 \hspace{0.2cm} 0.79 \hspace{0.2cm} 3.13\\
				\vspace{0.05cm} 
				
				[2~~~0~~~0]          & -  
				& 0.22 \hspace{0.2cm} 0.25 
				& 0.85 \hspace{0.2cm} 4.29 
				& 0.42 \hspace{0.2cm} 0.25 \hspace{0.2cm} 1.63  
				& 0.28 \hspace{0.2cm} 0.24 \hspace{0.2cm} 1.02 
				& 0.53 \hspace{0.2cm} 0.54 \hspace{0.2cm} 1.49\\

			\end{tabular}
		\end{ruledtabular}
		
		
		\label{Tab.1}
	\end{table*}

	The SRO is further investigated by means of CNA \cite{Faken1994} around each species for the binary and ternary liquid alloys. In Table \ref{Tab.1}, results are gathered for two temperatures showing the most important bonded pairs determined from IS configurations. Five-fold symmetry is represented by the 555 pairs as well as the 554 ones, which are a distorted version of 555 \cite{Faken1994}. A high degree of FFS characterizes the SRO of all alloys in the liquid at $T=1000$~K, and becomes even more pronounced in the undercooled state at $T=600$~K.
	
	As a general observation, for all systems, the FFS around Mg is higher than around Al, while for Si it is lower. For Mg, the presence of 666 pairs could indicate the presence of bcc ordering when associated to 444 pairs. However, the proportion of 444 is too small, suggesting therefore Frank-Kasper polyhedra, mainly Z14 and Z15, to be more likely and consistent with a coordination number close to 15. For Si, the presence of 421 and 422 pairs is an indication of a significant cubic symmetry. The respective specific SRO around Mg and Si is reinforced in the undercooled regime. The 433 pairs can be associated either to very distorted 555 pairs or to distorted cubic symmetry. It generally decreases with decreasing temperature, indicating that the liquid becomes more structured. For both temperatures, the addition of Mg or Si atoms leads to a global increase of the FFS in each alloys by up to 10\%, and sometimes even more around Al atoms with respect to pure liquid Al \cite{Jakse2013}. The increase of FFS around Al in alloys is consistent with a higher compacity, as mentioned above \cite{Frank1952}. Table \ref{Tab.1} also reports the presence of a small amount of 300 and 200 pairs in ternary alloys, also found in the Mg$_2$Si phase (see Table S3 in the SI file) and consistently with the typical distances found in the partial-pair correlation functions and bond-angle distributions. Nevertheless, it is difficult a this stage to claim the presence of precursors of the Mg$_2$Si phase that might form during slow solidification \cite{Liu1999}, given the small amount of these pairs present within the short time span of the simulations. This would require to perform large-scale MD simulations of the homogeneous nucleation process with a retrained ML potential \cite{Jain2021}, which is beyond the scope of the present paper.
	
	Interestingly, addition of Mg in pure Al to form Al$_{90}$Mg$_{10}$ or Si to form Al$_{90}$Si$_{10}$ leads respectively to a global increase or decrease of ISRO. Then, adding Si in Al$_{90}$Mg$_{10}$ or adding Mg in Al$_{90}$Si$_{10}$ to form the ternary Al$_{80}$Mg$_{10}$Si$_{10}$ alloy follows the same trend, namely decreasing or increasing the FFS of the initial corresponding binary alloy. Finally, in the ternary alloys, the same rule holds again, as the FFS is favored globally and around Al when the Mg content is increased at constant Si composition and disfavored when Si content is increased at constant Mg composition. Therefore, modifying the respective compositions of Mg and Si affects the local order in Al-Mg-Si alloy. More particularly, adapting ISRO might impact diffusion \cite{Pasturel2017}, thereby contributing to control solidification morphology.


	\subsection{Structure-diffusion relationship}
	
\begin{figure}[t!]
		\begin{center}
			\includegraphics[angle=0 ,width=0.6\textwidth]{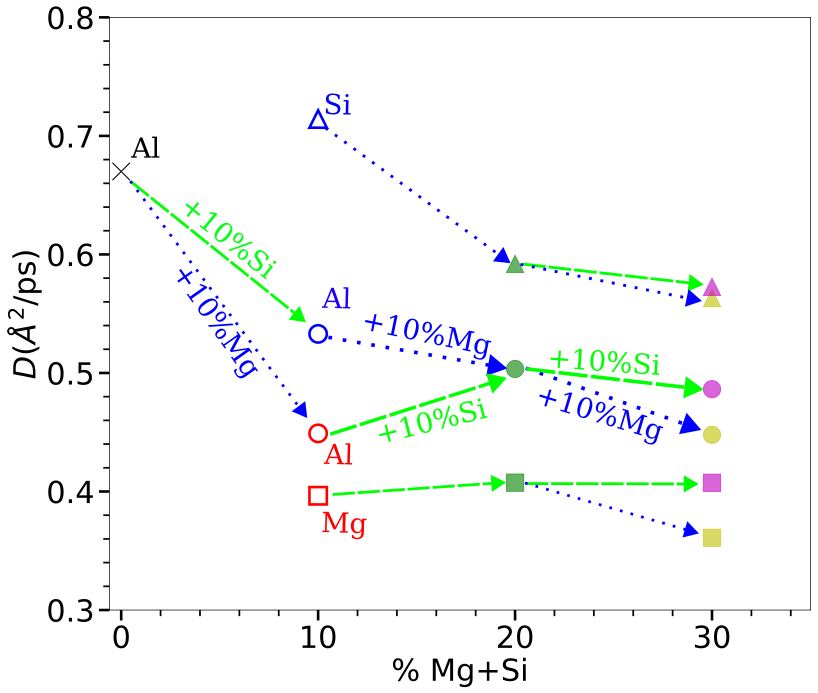} 
		\end{center}
		\caption{Diffusion coefficients for all the liquid alloys as function of composition at $T=$1000K. The open symbols represent the binary liquid alloys (Red: Al$_{90}$Mg$_{10}$; Blue: Al$_{90}$Si$_{10}$) and the closed symbols represent ternary liquid alloys (Green: Al$_{80}$Mg$_{10}$Si$_{10}$; Dark yellow: Al$_{70}$Mg$_{20}$Si$_{10}$; Magenta: Al$_{70}$Mg$_{10}$Si$_{20}$). Similar tendencies of diffusion coefficients have been found for other undercooling temperatures (see Figs. S29 and S30 in the SI file). The cross represent the value for pure liquid Al from \cite{Jakse2023,Jakse2013}  
		}	
		\label{fig4}   
	\end{figure}

	In order to determine self-diffusion coefficients for all the studied alloys, the mean-square-displacement $R^2_i(t)$ (Eq. \ref{eq.3}) is calculated for each species $i$ on the production stage of the simulations after an equilibration of 30 ps. As can be easily seen on the log-log plot of the $R^2_i(t)$ curves in Figs. S24-S28 in the SI file, this physical quantity is divided into two main parts, starting at small times by the ballistic motions and followed by a linear diffusive regime roughly after $t=$ 20 ps for all alloys. In time spans between the two regimes, a slowing down of the MSD can be seen, corresponding to the cage effects, which reinforces with decreasing temperature. The self diffusion coefficient is obtained by a linear regression of $R^2_i(t)$ in the long-time diffusive part and by using Eq. \ref{eq.4}. \\
	\indent Results of diffusion coefficients are drawn as a function of composition at $T = 1000$ K
	(see Fig. \ref{fig4}). Starting from pure liquid Al, which was studied in a previous contribution \cite{Jakse2013}, binary liquid alloys are obtained by adding Mg or Si respectively for Al$_{90}$Mg$_{10}$ and Al$_{90}$Si$_{10}$, leading to a downshift of diffusion coefficients of Al for both alloys. From the structural analysis in the preceding section, the enhancement of the compacity around Al atoms together with the more pronounced ISRO might be at the origin of this slowing down for the two binary alloys. Nevertheless, Fig. \ref{fig4} clearly shows that adding Mg or Si in liquid Al does not
	yield to the same consequences.
	ISRO is more pronounced in Al$_{90}$Mg$_{10}$ than in Al$_{90}$Si$_{10}$ leading to lower diffusion of Al atoms. Moreover, in Al$_{90}$Mg$_{10}$ the higher icosahedral ordering around Mg atoms leads to an even lower value of $D_{\rm Mg}$. 
	For Al$_{90}$Si$_{10}$, the local ordering is rather cubic and less compact around Si giving rise to higher Si diffusion. It is worth
	mentioning that from AIMD simulation studies, Si is known as a fast diffuser \cite{Jakse2009}, giving
	rise to a more important decoupling in the binary alloys. For both binaries, a decoupling of
	the self-diffusion coefficients is observed, which is more pronounced in Al$_{90}$Si$_{10}$, and might
	be a consequence of structural heterogeneities, as revealed by the CNA analysis. \\
	\indent The Al$_{80}$Mg$_{10}$Si$_{10}$ ternary alloy can be formed, as mentioned above, either by adding
	Si in Al$_{90}$Mg$_{10}$ or adding Mg in Al$_{90}$Si$_{10}$, in substitution of Al atoms. Fig. \ref{fig4} shows that
	both paths are consistent with the mechanism unveiled above in which the increase of Mg
	content promotes ISRO and slows down the diffusion of each species, while the increase of Si
	composition induces cubic local ordering around Si and disfavors ISRO with the consequence
	of enhancing the diffusion of the species. This is all the more true for Al$_{70}$Mg$_{20}$Si$_{10}$ and
	Al$_{70}$Mg$_{10}$Si$_{20}$ obtained from Al$_{80}$Mg$_{10}$Si$_{10}$, respectively by increasing the Mg content at constant Si composition and increasing Si content and maintaining the Mg composition.
	
	\section{Conclusion}
	
	In conclusion, AIMD simulations were conducted in the liquid and undercooled states
	for Al-rich Al-Mg-Si ternary system, prototypical of 6xxx alloys, together with their binary
	counterparts, namely Al-Mg and Al-Si, with the goal of understanding the relationship
	between structure and diffusion. Our results show that the presence of Mg and/or Si leads to
	a local structure around Al significantly more compact in the alloys than in pure Al, which
	has the consequence of a global lowering of the diffusion. Moreover, an opposite influence of
	Mg or Si addition in the alloys on the liquid structure and dynamics was found. Whereas
	the presence of Mg in the alloys promotes ISRO, Si atoms show a tendency to disfavor
	ISRO and to form a cubic local ordering. These effects are shown to impact the diffusion of
	all species depending on the respective Mg and Si contents, and suggest the mechanism by
	which an increase of Mg content generally slows down the diffusion of each species, while
	the increase of Si composition enhances their diffusion. This demonstrates that chemical
	short-range order (CSRO) between Mg and Si is able to neutralize the ISRO-killing effect of Si atoms without annihilating the ISRO-promoting effect of Mg atoms. The evidence of such
	an interplay between topological and chemical short range orders provides significant insight
	into alloy designing principles. It may constitute the basis of any future investigation of the
	effect of other additional elements to the structural-dynamic properties of this system, with
	the possibility to fine tune the compositions to control diffusion mechanisms and thus the
	solidification process.
	
	\section*{Acknowledgement}
	
	We acknowledge the CINES and IDRIS under Project No. INP2227/72914, as well as
	CIMENT/GRICAD for computational resources. This work was performed within the
	framework of the Centre of Excellence of Multifunctional Architectured Materials CEMAM-
	ANR-10-LABX-44-01 funded by the "Investments for the Future" Program. This work has
	been partially supported by MIAI@Grenoble Alpes (ANR-19-P3IA-0003) and international SOLIMAT projet ANR-22-CE92-0079-01. Discussions within
	the French collaborative network in high-temperature thermodynamics GDR CNRS3584
	(TherMatHT) and in artificial intelligence in materials science GDR CNRS 2123 (IAMAT)
	are also acknowledged.
	
	\bibliographystyle{apsrev}
	\normalem
	\bibliography{References.bib}
	
	\newpage
	
	\renewcommand{\thefigure}{S\arabic{figure}}
	\renewcommand{\thetable}{S\Roman{table}}
	\renewcommand{\theequation}{S\arabic{equation}}
	
	\begin{center}
		{\Large Supplemental Material}
	\end{center}

\begin{table}[h]
	\begin{center}
		\centering
		
		\begin{tabular}{cccccc}	
			\hline		
			& Al$_{90}$Mg$_{10}$ & Al$_{90}$Si$_{10}$ & Al$_{80}$Mg$_{10}$Si$_{10}$ & Al$_{70}$Mg$_{20}$Si$_{10}$& Al$_{70}$Mg$_{10}$Si$_{20}$\\
			\hline
			$N_{\rm Al}$	 &  230 & 230 & 204 & 179 & 179 \\
			$N_{\rm Mg}$   &  26  &  -  & 26  & 51  & 26  \\
			$N_{\rm Si}$   &   -  & 26  & 26  & 26  & 51  \\
			\hline
			
		\end{tabular}
	\end{center}
	\caption{Number of atoms of each species for all liquid alloys used in VASP simulations. The total number of atoms is $N=256$.}
	\label{Tab.1}
\end{table}

\begin{figure}[h]
	\begin{center}
		\includegraphics[angle=0 ,width=1\textwidth]{ 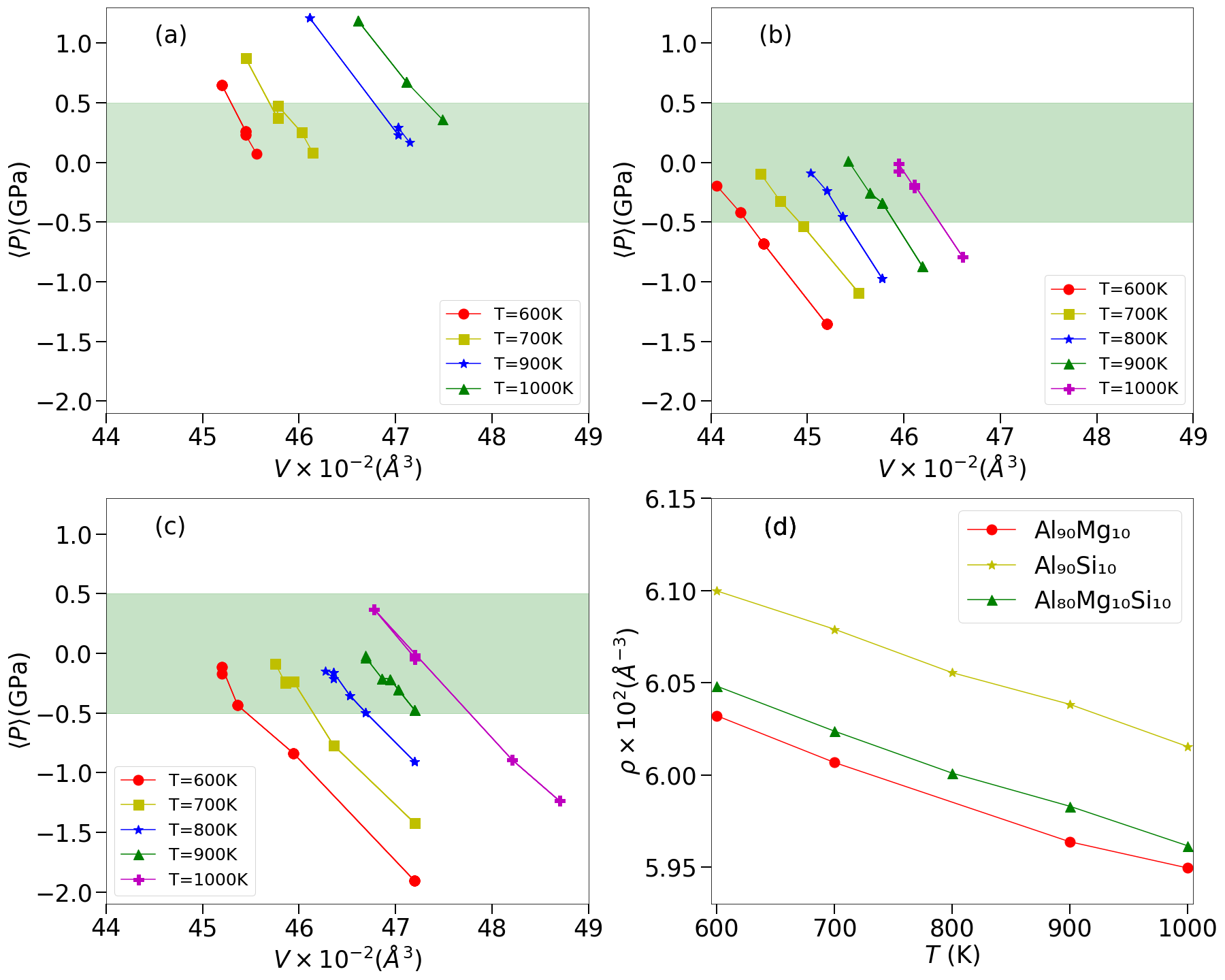} \hfill	
	\end{center}
	\caption{Mean pressure $\left< P \right>$ as function of the volume $V$ of the simulation box, calculated at different temperatures for Al$_{90}$Mg$_{10}$ (a), Al$_{90}$Si$_{10}$ (b), Al$_{80}$Mg$_{10}$Si$_{10}$ (c) liquid alloys. The green horizontal background corresponds to the region where $\left< P \right>$ converges to a negligible value. Interpolation or slight extrapolation of the volume to zero pressure was done to fix the density, which is represented in (d). Using these densities, simulations are performed to determine structural and dynamic properties. For the Al$_{70}$Mg$_{20}$Si$_{10}$, the optimal density is $\rho=0.0528$ \AA$^{-3}$ and $\rho=0.0551$ \AA$^{-3}$ respectively at $T = 1000$ K and $T = 600$ K. For the Al$_{70}$Mg$_{10}$Si$_{20}$ liquid alloy, $\rho^{T=1000 \rm K}=0.0556$ \AA$^{-3}$ and $\rho^{T = 600 \rm K}=0.0575$ \AA$^{-3}$.}
	
\end{figure}

\begin{figure}[h]
	\begin{center}
		\includegraphics[angle=0 ,width=1\textwidth]{ 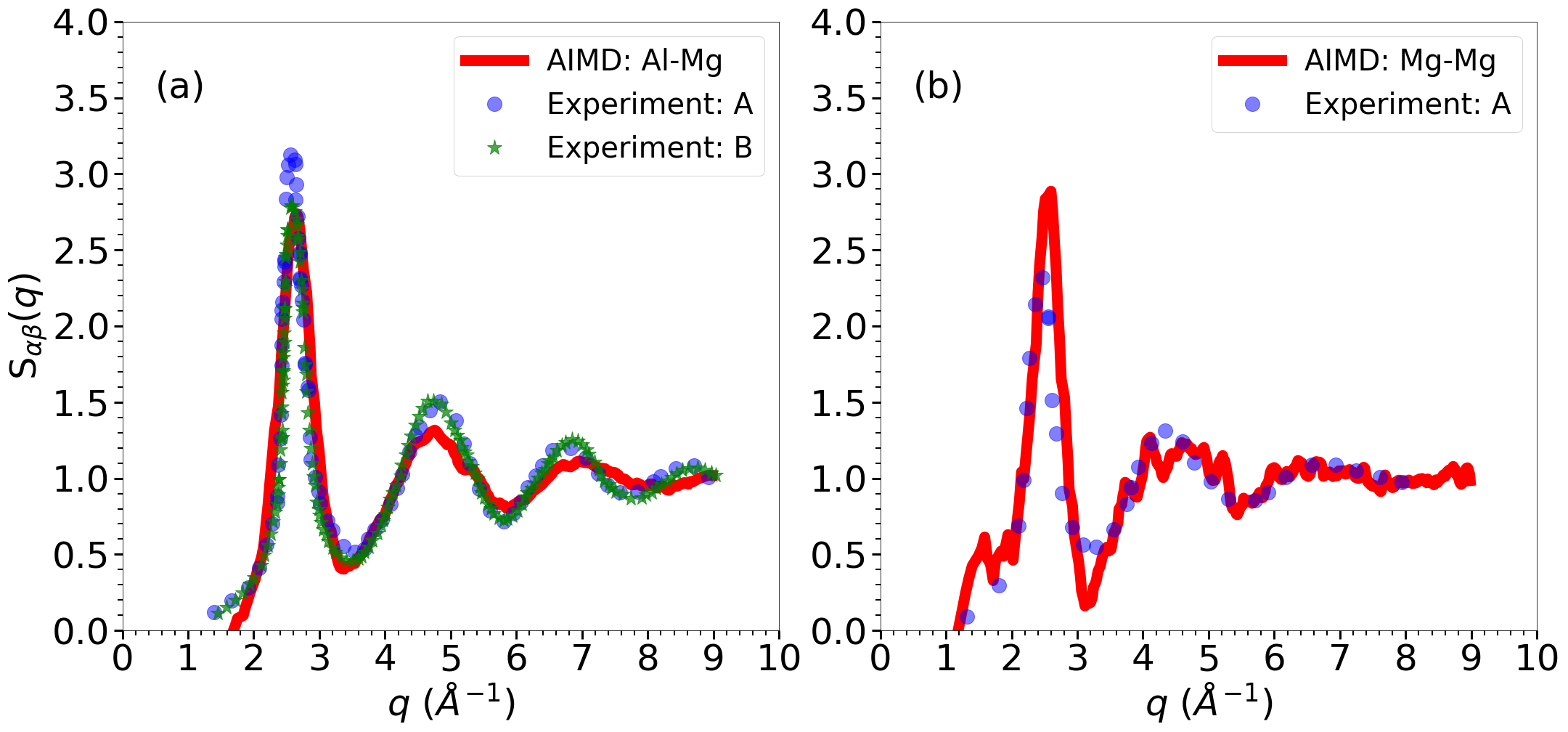} \hfill	
	\end{center}
	\caption{AIMD calculations (red curves) of Al-Mg (a) and Mg-Mg (b) partial structural factors in the Al$_{90}$Mg$_{10}$ liquid alloy. These calculations were done in the framework of Faber-Ziman formalism \cite{Faber1965}. Experimental X-ray diffraction data \cite{Waseda1973} are compared to the AIMD calculations. Experiment A (blue circles) and B (green stars) refers to two experimental methods used by Waseda \textit{et al.} \cite{Waseda1973} for assessing the partial structural factor of the Al-Mg pair. It appears that Method B is closer to the AIMD calculations. Experimental data using Method A for Mg-Mg pair are very well reproduced by AIMD calculations. No data were reported for the Mg-Mg pair using Method B.}
	
\end{figure}

\begin{figure}[h]
	\begin{center}
		\includegraphics[angle=0 ,width=0.49\textwidth]{ 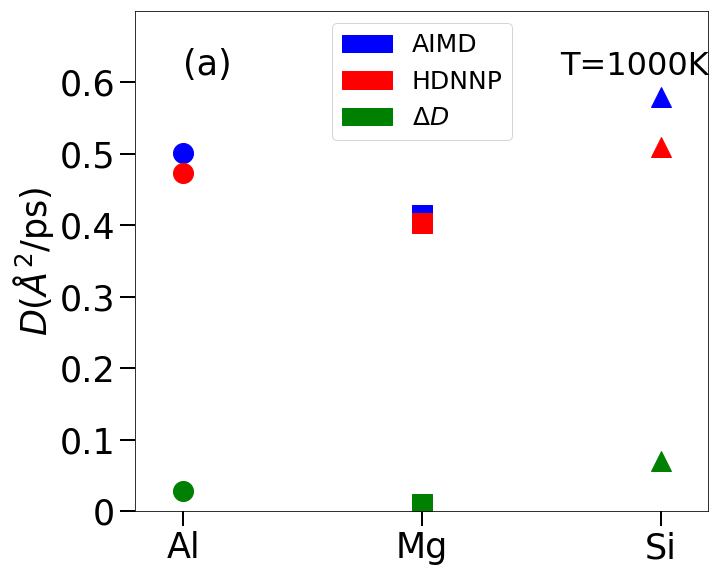}
		\includegraphics[angle=0 ,width=0.49\textwidth]{ 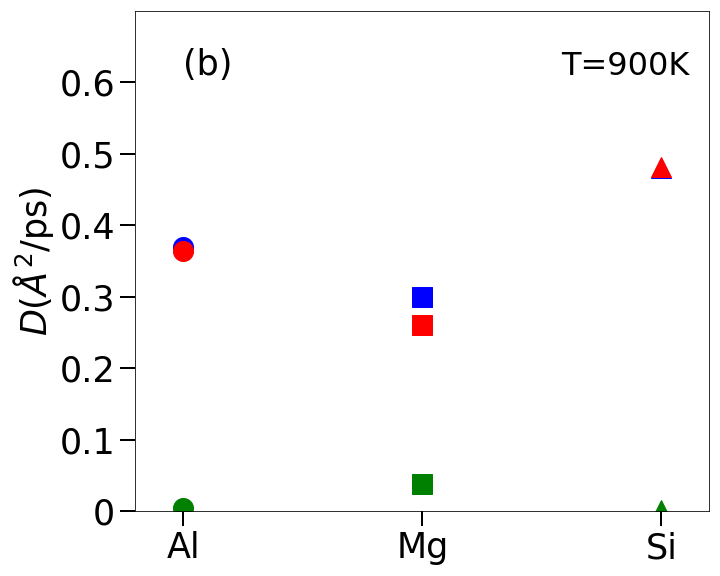} 
		\includegraphics[angle=0 ,width=0.49\textwidth]{ 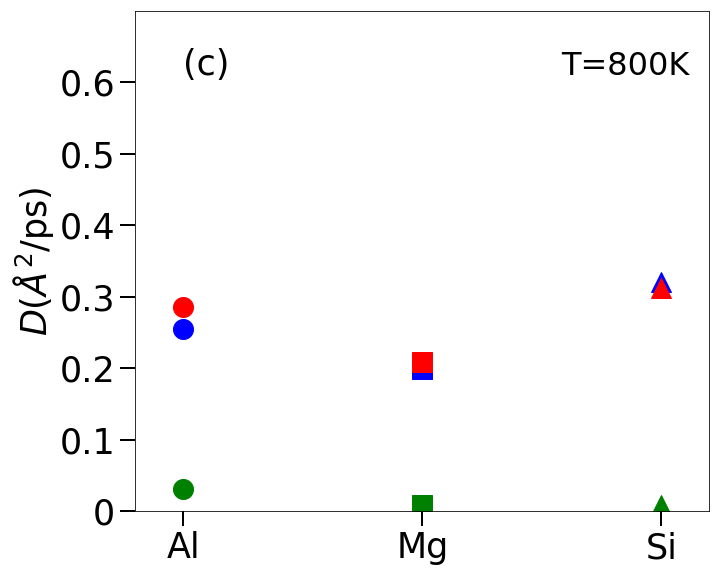} 
		\includegraphics[angle=0 ,width=0.49\textwidth]{ 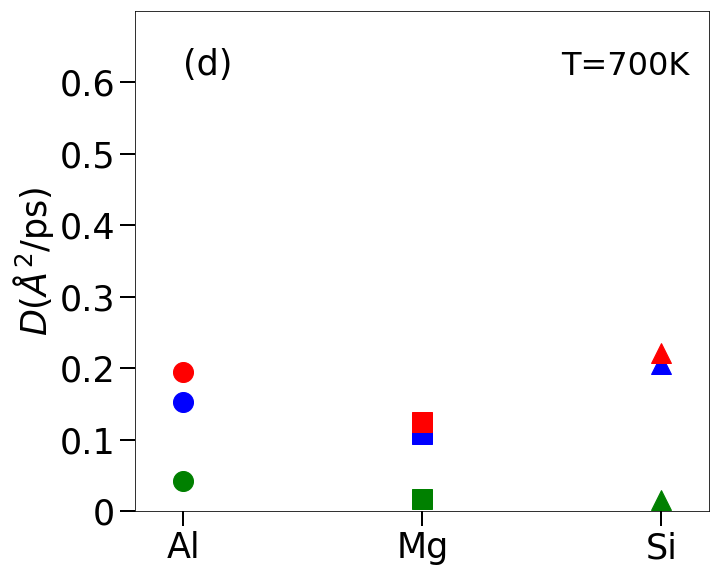} 
		\includegraphics[angle=0 ,width=0.49\textwidth]{ 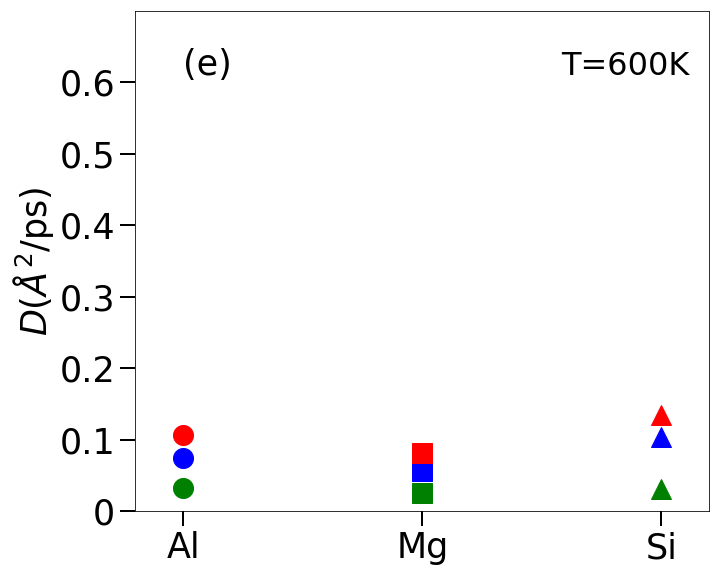}

	\end{center}
	\caption{Diffusion coefficients for all the elements in the ternary Al$_{80}$Mg$_{10}$Si$_{10}$ liquid alloy, calculated at different temperatures using AIMD and HDNNP methods. $\Delta D$ represents the absolute difference of diffusion coefficients of the two methods.
	}
	
\end{figure}

\begin{figure}[h]
	\begin{center}
		\includegraphics[angle=0 ,width=0.49\textwidth]{ 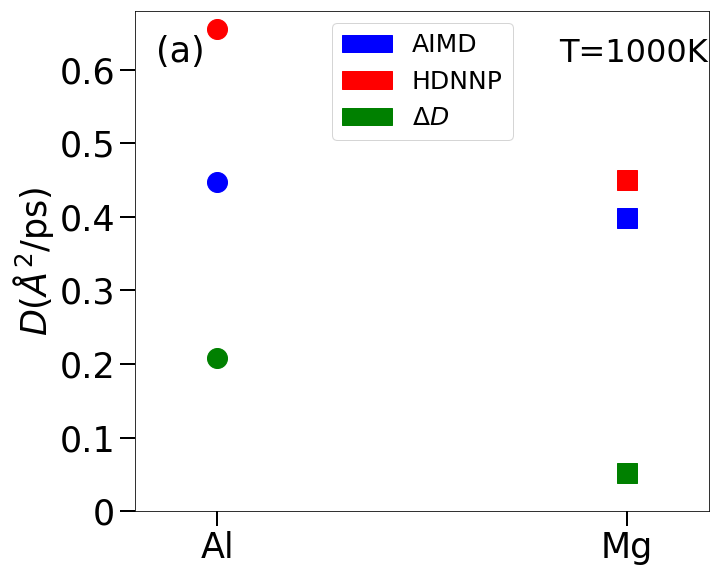} 
		\includegraphics[angle=0 ,width=0.49\textwidth]{ 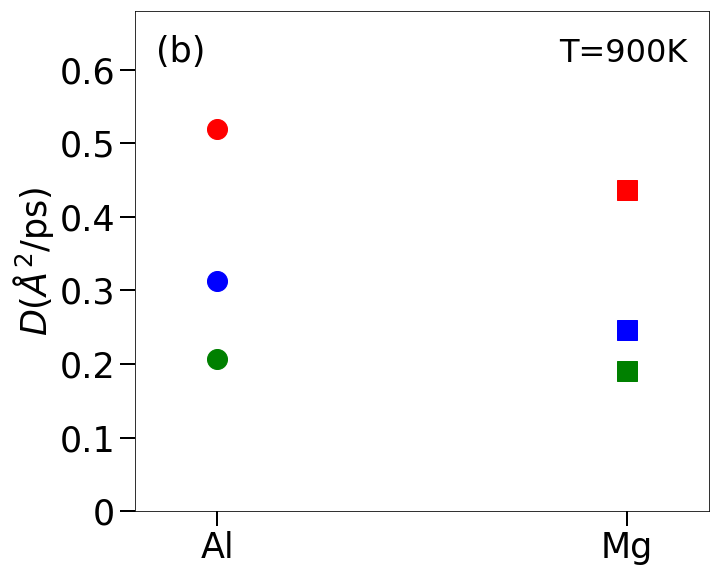} 
		\includegraphics[angle=0 ,width=0.49\textwidth]{ 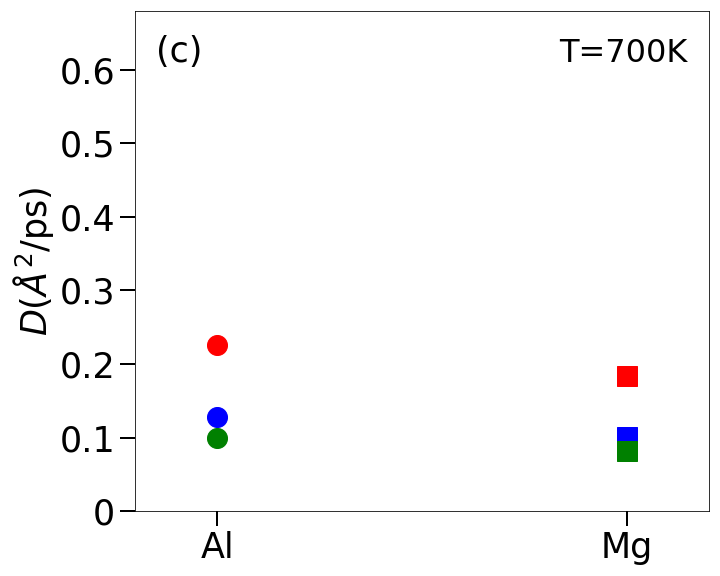} 
		\includegraphics[angle=0 ,width=0.49\textwidth]{ 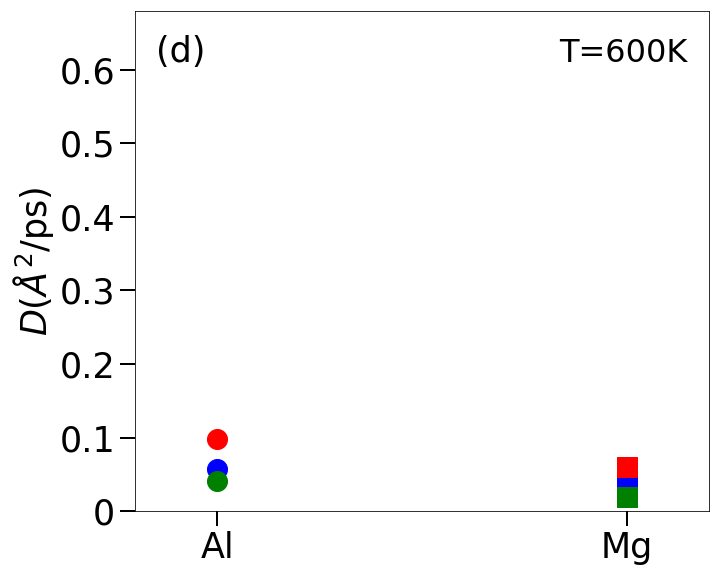}

	\end{center}
	\caption{Diffusion coefficients for all the elements in the binary Al$_{90}$Mg$_{10}$ liquid alloy, calculated at different temperatures using AIMD and HDNNP methods. $\Delta D$ represents the absolute difference of diffusion coefficients of the two methods.
	}
	
\end{figure}

\begin{figure}[h]
	\begin{center}
		\includegraphics[angle=0 ,width=0.95\textwidth]{ 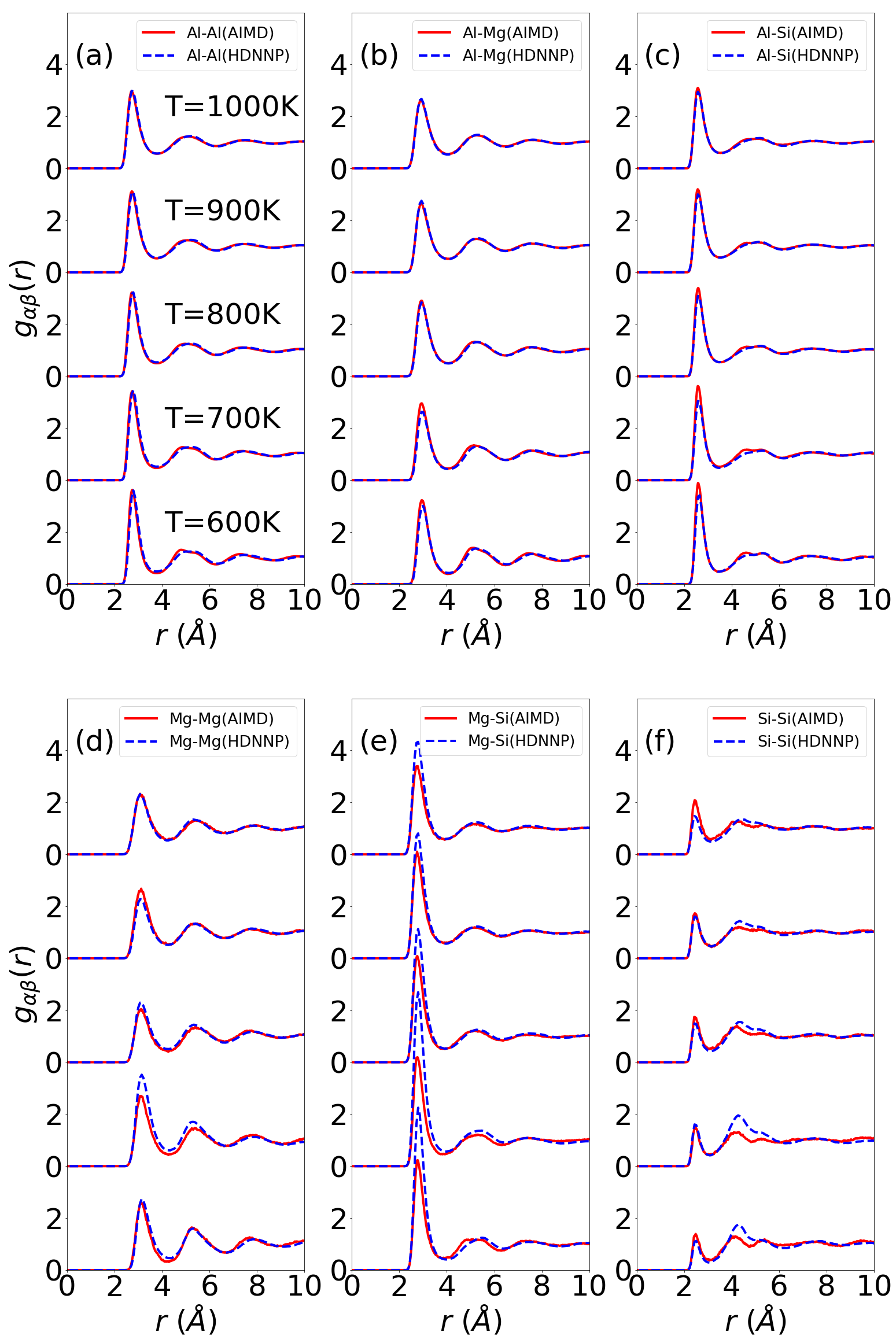} 	
	\end{center}
	\caption{Comparison of partial pair-correlation functions of the ternary Al$_{80}$Mg$_{10}$Si$_{10}$ liquid alloy using HDNNP and AIMD methods.}
	
\end{figure}

\begin{figure}[h]
	\begin{center}
		\includegraphics[angle=0 ,width=1.\textwidth]{ 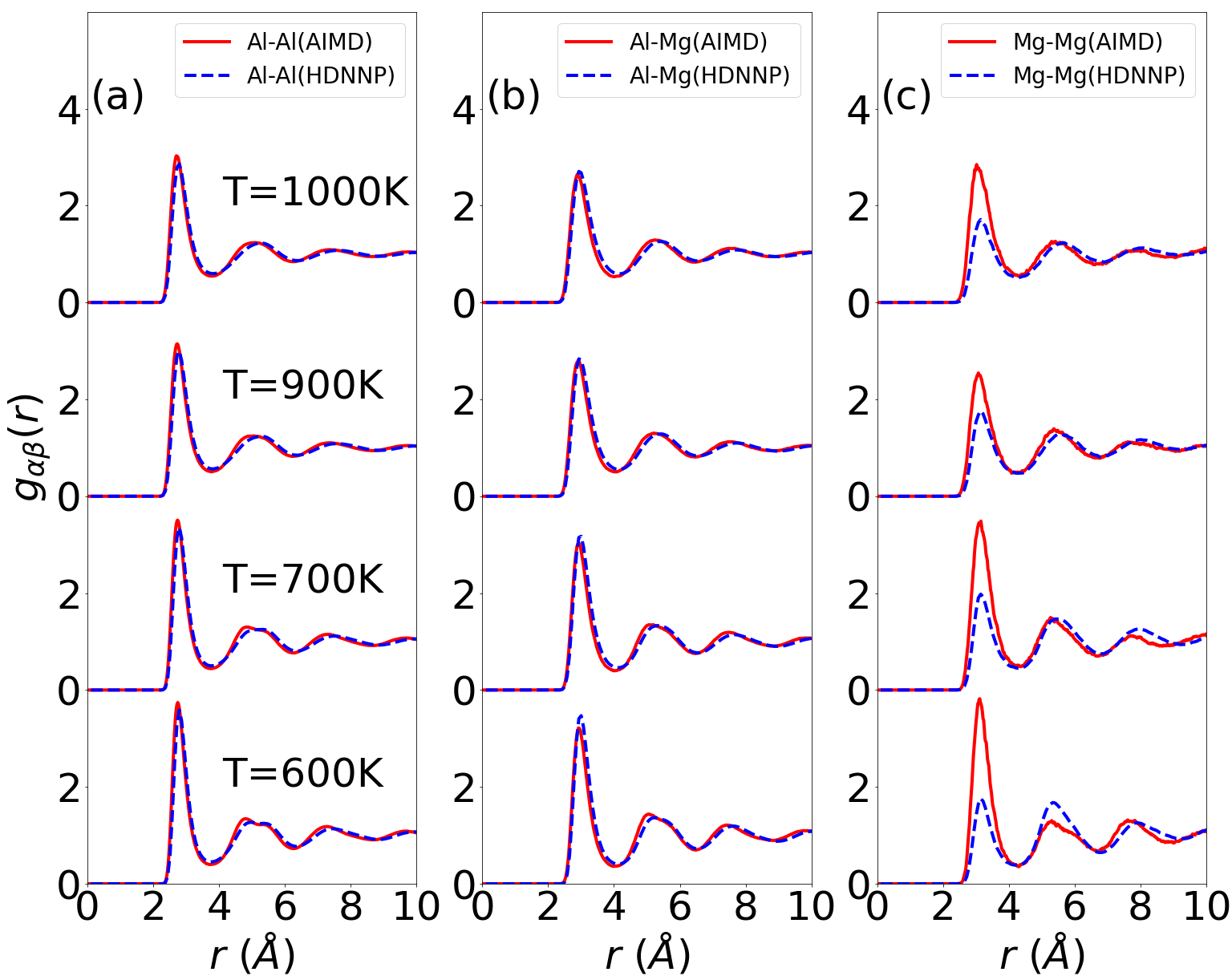} 	
	\end{center}
	\caption{Comparison of partial pair-correlation functions of the binary Al$_{80}$Mg$_{10}$ liquid alloy using HDNNP and AIMD methods.}
	
\end{figure}

\begin{figure}[h]
	\begin{center}
		\includegraphics[angle=0 ,width=1\textwidth]{ 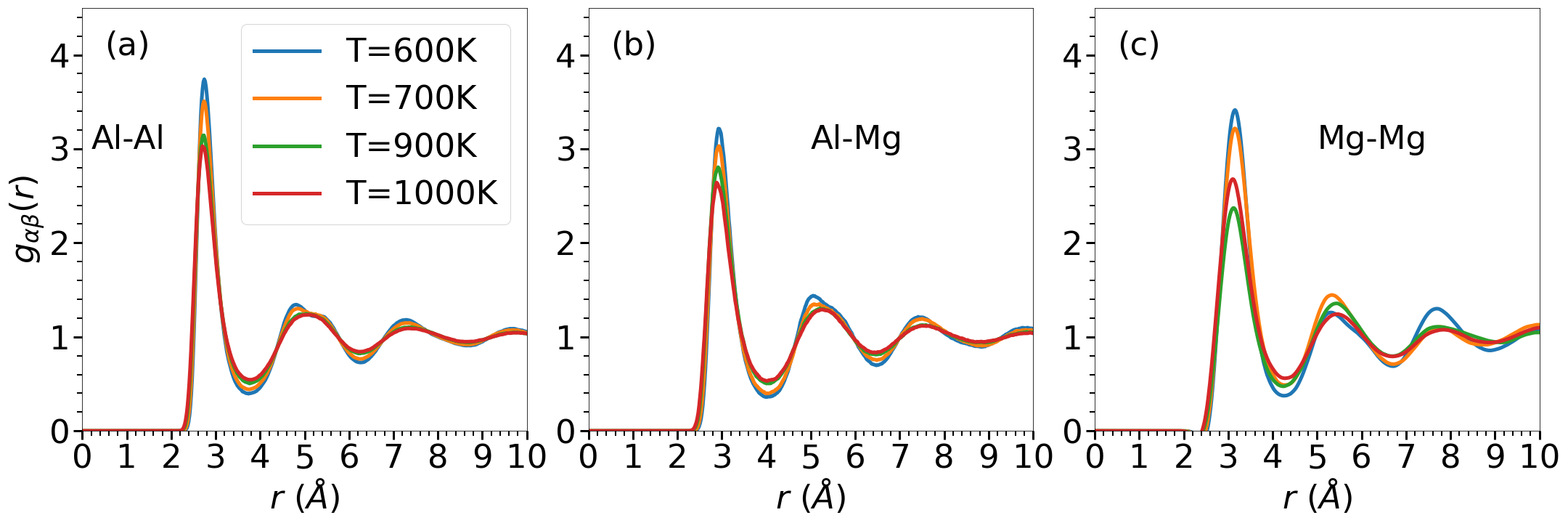} 	
	\end{center}
	\caption{Partial pair-correlation function of the Al$_{90}$Mg$_{10}$ liquid alloy from AIMD simulations at different temperatures, as function of the distance $r$ from a reference atom.}
	
\end{figure}	

\begin{figure}[h]
	\begin{center}
		\includegraphics[angle=0 ,width=1\textwidth]{ 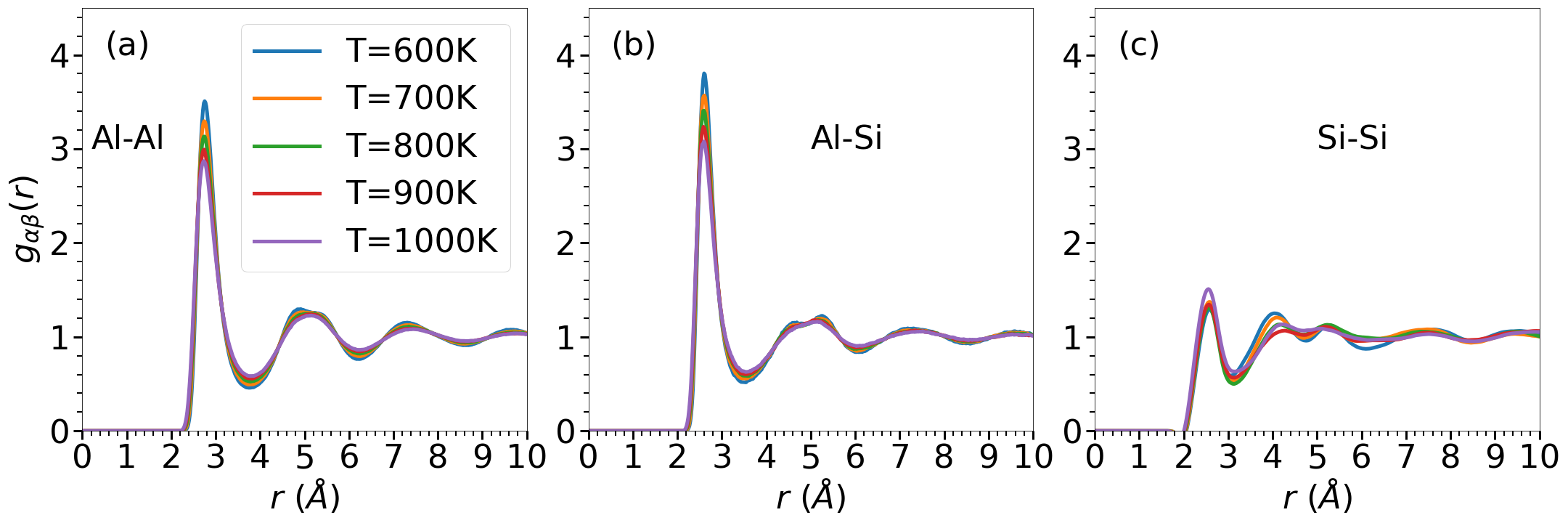} 	
	\end{center}
	\caption{Partial pair-correlation function of the Al$_{90}$Si$_{10}$ liquid alloy from AIMD simulations at different temperatures.}
	
\end{figure}	

		
	

\begin{figure}[!h]
	\begin{center}
		\includegraphics[angle=0 ,width=1\textwidth]{ 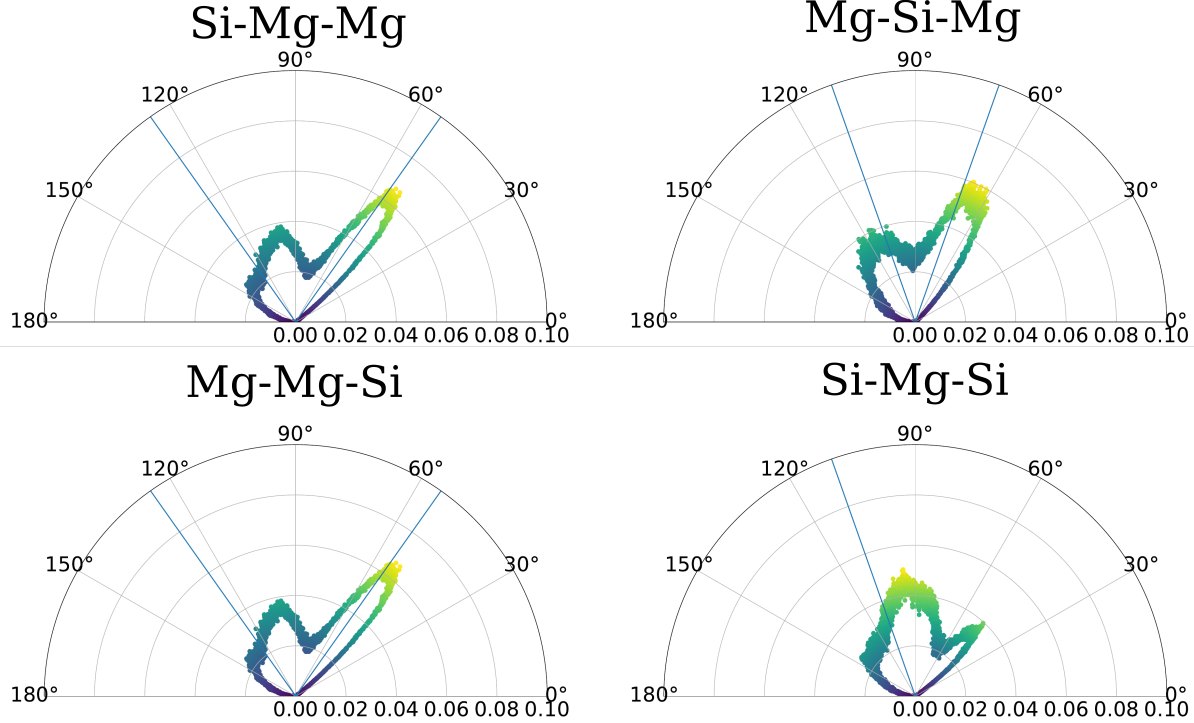} 
		
	\end{center}
	\caption{Bond angle distribution in liquid Al$_{80}$Mg$_{10}$Si$_{10}$ alloy (distribution of angles) and in Mg$_2$Si solid crystal (straight lines). }
	
\end{figure}

\begin{figure}[!h]
	\begin{center}
		\includegraphics[angle=0 ,width=0.8\textwidth]{ 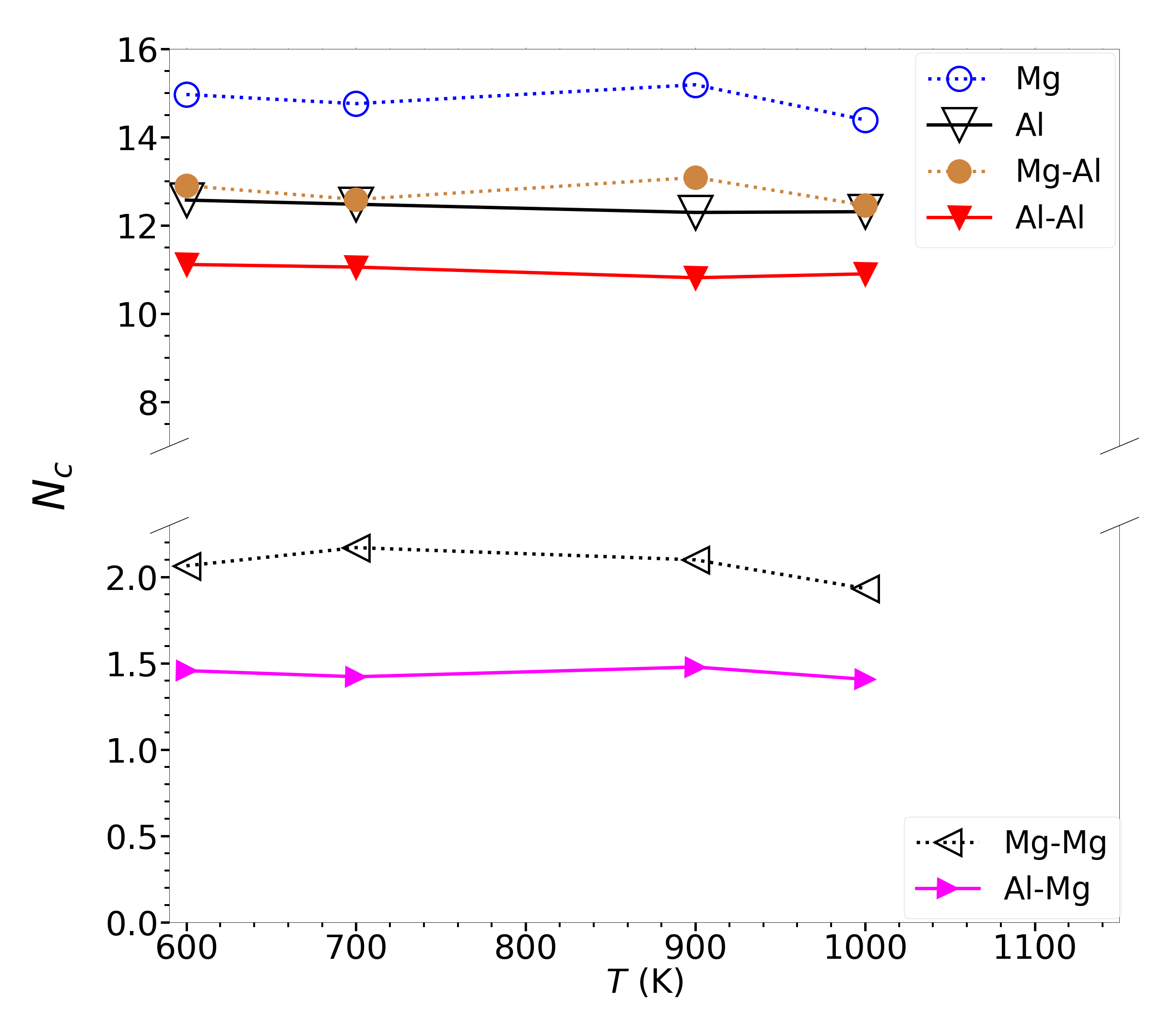} 	
		
		\includegraphics[angle=0 ,width=0.8\textwidth]{ 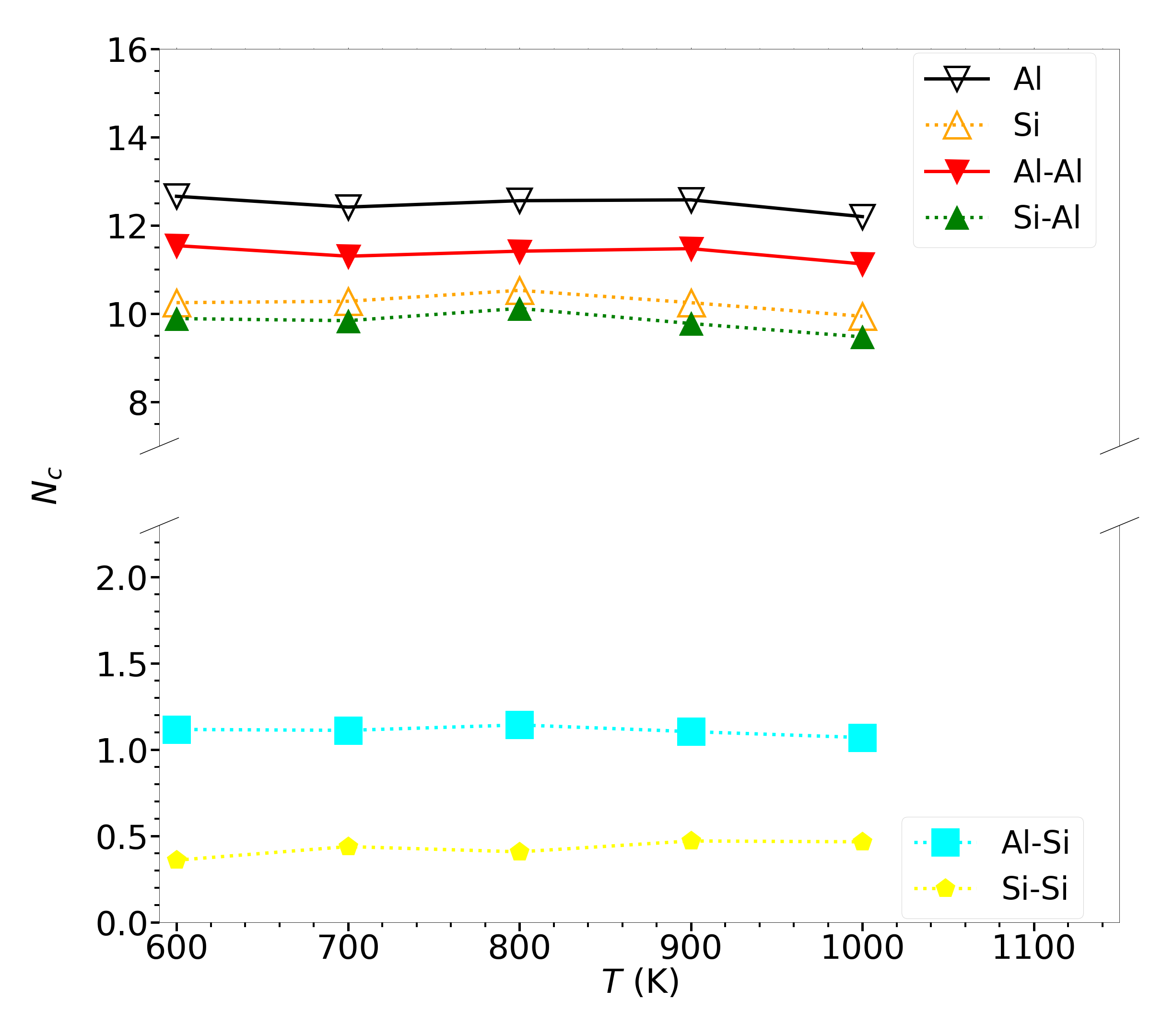}
		
	\end{center}
	\caption{Total and partial coordination numbers as function of temperature in the two binary liquid alloys: Al$_{90}$Mg$_{10}$ (upper panel) and Al$_{90}$Si$_{10}$ (lower panel).}
	
\end{figure}


\begin{table}[h]

	\begin{tabular}{ccccccc}

		\hline 	
		$r_{\rm max}$ (\AA) 
		&    Al \cite{Jakse2013}        
		&    Al$_{90}$Mg$_{10}$    
		&    Al$_{90}$Si$_{10}$     
		&    Al$_{80}$Mg$_{10}$Si$_{10}$     
		&    Al$_{70}$Mg$_{20}$Si$_{10}$    
		&    Al$_{70}$Mg$_{10}$Si$_{20}$  \\ 
		\hline
		Al-Al  &    2.7      &    2.7    &     2.7     &      2.7      &      2.7      &      2.7  \\
		Al-Mg  &    -        &    2.9    &      -      &      2.9      &      2.9      &      2.9  \\
		Al-Si  &    -        &     -     &     2.6     &      2.6      &      2.6      &      2.6  \\
		Mg-Mg  &    -        &    3.1    &      -      &      3.1      &      3.1      &      3.1   \\
		Mg-Si  &    -        &     -     &      -      &      2.7      &      2.7      &      2.7   \\
		Si-Si  &    -        &     -     &     2.5     &      2.5      &      2.5      &      2.5   \\
		
		\hline
		
	\end{tabular}
	\caption{Distances at the first maximum of $g_{\alpha\beta}(r)$ curves ($r_{\rm max}$) calculated at $T=$ 1000 K for five Al-based liquid alloys. Results for pure liquid Al, reported in Ref. \cite{Jakse2013}, are also shown for comparison.}
	\label{Tab.2}
\end{table}

\begin{table}[!h]

	\begin{tabular}{cccccccccc}
		
		
		Reference
		& mp-134
		& mp-110 
		& mp-153
		& mp-1194114
		& mp-1367
		& mp-2151\\
		\hline
		
		& Al
		& Mg 
		& Mg
		& Mg$_{13}$Al$_{16}$
		& Mg$_{2}$Si
		& Mg$_{17}$Al$_{12}$\\
		
		\hline
		
		CNA (\%) 
		&            
		&            
		&            
		& Mg$_{13}$ \hspace{0.2cm} Al$_{16}$ 
		& Mg$_{2}$  \hspace{0.2cm} Si 
		& Mg$_{17}$ \hspace{0.2cm} Al$_{12}$\\
		\hline

		$[5~5~5] + [5~4~4]$  & 0.00  
		& 0.00  
		& 0.00  
		& 7.50 \hspace{0.2cm} 13.8         
		& 0.00 \hspace{0.2cm} 0.00         
		& 6.00 \hspace{0.2cm} 18.0 \\      
		
		$[6~~~6~~~6]$        & 0.00   
		& 57.0  
		& 0.00   
		& 2.50 \hspace{0.2cm} 2.30  
		& 0.00 \hspace{0.2cm} 0.00      
		& 4.00 \hspace{0.2cm} 0.00 \\   
		
		
		$[4~~~4~~~4]$        & 0.00  
		& 43.0 
		& 0.00 
		& 0.00 \hspace{0.2cm} 0.00  
		& 0.00 \hspace{0.2cm} 0.00   
		& 0.00 \hspace{0.2cm} 0.00 \\ 
		
		$[4~~~3~~~3]$        & 0.00  
		& 0.00  
		& 0.00  
		& 45.0 \hspace{0.2cm} 27.9  
		& 0.00 \hspace{0.2cm} 0.00  
		& 48.0 \hspace{0.2cm} 18.1 \\ 
		
		$[4~~~2~~~2]$        & 0.00  
		& 0.00 
		& 50.0 
		& 0.00 \hspace{0.2cm} 13.9  
		& 0.00 \hspace{0.2cm} 0.00  
		& 0.00 \hspace{0.2cm} 18.1 \\ 
		
		
		$[4~~~2~~~1]$        & 100  
		& 0.00 
		& 50.0 
		& 22.0 \hspace{0.2cm} 20.9  
		& 0.00 \hspace{0.2cm} 0.00  
		& 18.0 \hspace{0.2cm} 27.2 \\ 
		
		
		

		
		[3~~~1~~~1]          & 0.00  
		& 0.00  
		& 0.00 
		& 15.0 \hspace{0.2cm} 13.9  
		& 0.00 \hspace{0.2cm} 0.00  
		& 12.0 \hspace{0.2cm} 18.1 \\ 
		
		[3~~~0~~~0]          & 0.00  
		& 0.00  
		& 0.00 
		& 0.00 \hspace{0.2cm} 0.00
		& 40.0 \hspace{0.2cm} 100  
		& 0.00 \hspace{0.2cm} 0.00 \\ 
		
		[2~~~1~~~1]          & 0.00  
		& 0.00  
		& 0.00 
		& 7.50 \hspace{0.2cm} 6.90  
		& 0.00 \hspace{0.2cm} 0.00  
		& 12.0 \hspace{0.2cm} 0.00 \\ 
		
		[2~~~0~~~0]          & 0.00  
		& 0.00  
		& 0.00 
		& 0.00 \hspace{0.2cm} 0.00  
		& 60.0 \hspace{0.2cm} 0.00  
		& 0.00 \hspace{0.2cm} 0.00 \\ 
		

		\hline 	
		
	\end{tabular}
	\caption{Abundance of pairs results obtained for different crystal phases retrieved from the Materials Project \cite{Materialsproject}.}
	\label{Tab.2}
\end{table}

\begin{figure}[h]
\begin{center}
	\includegraphics[angle=0 ,width=1\textwidth]{ 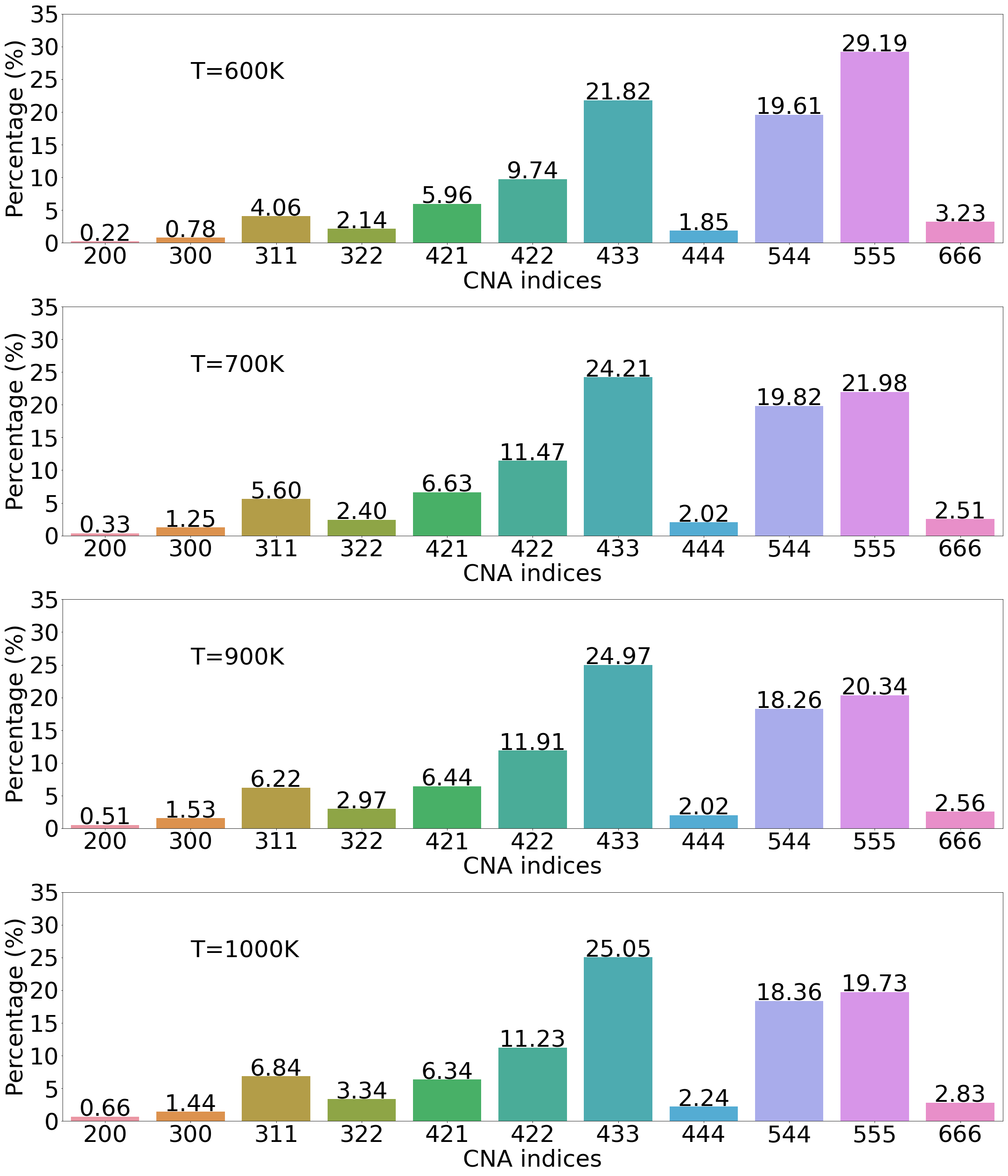} \hfill	
\end{center}
\caption{Abundance of pairs from the CNA centred on Al atoms in the Al$_{90}$Mg$_{10}$ binary liquid alloy at different temperatures.}

\end{figure}	

\begin{figure}[h]
\begin{center}
	\includegraphics[angle=0 ,width=1\textwidth]{ 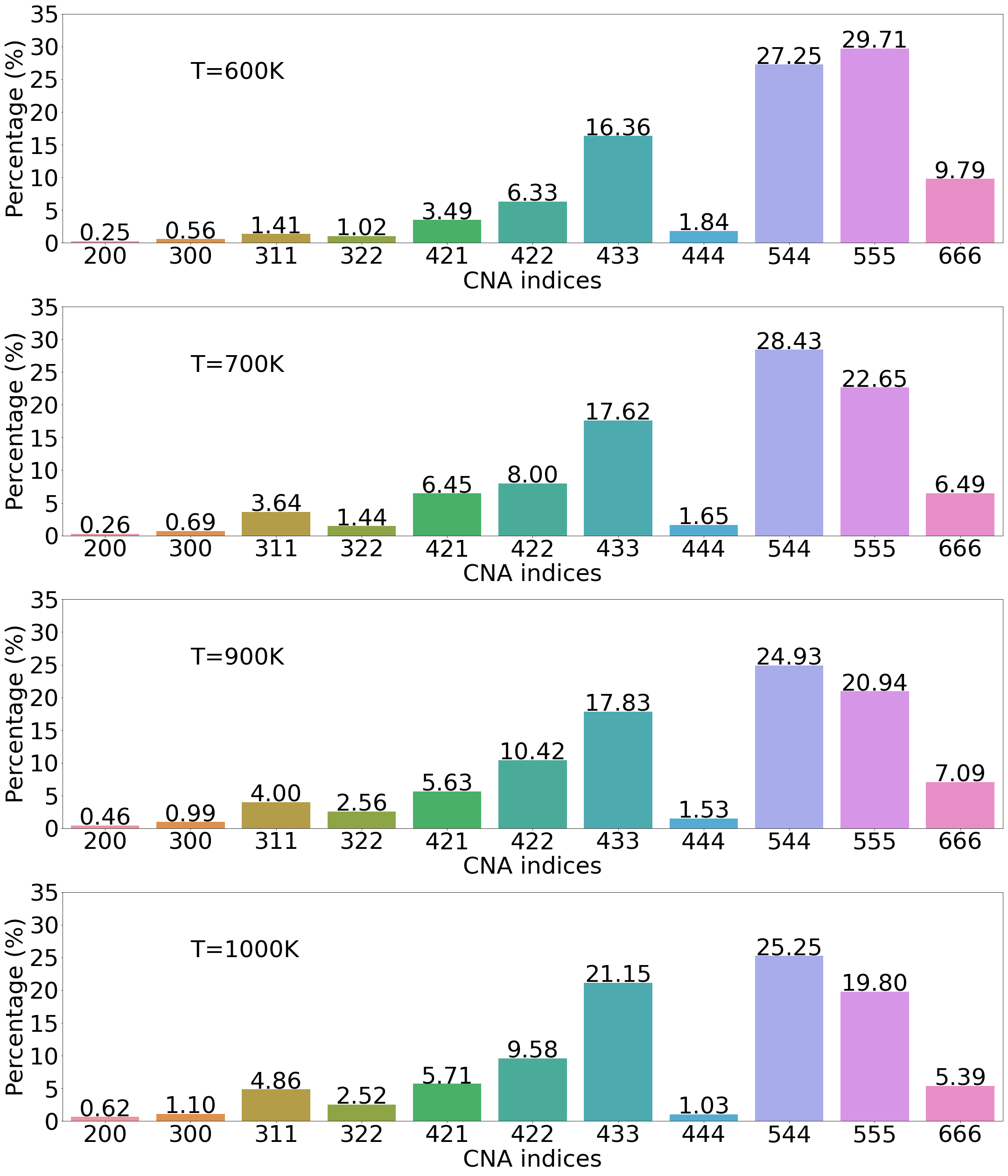} \hfill	
\end{center}
\caption{Abundance of pairs from the CNA centred on Mg atoms in the Al$_{90}$Mg$_{10}$ binary liquid alloy at different temperatures.}

\end{figure}

\begin{figure}[h]
	\begin{center}
		\includegraphics[angle=0 ,width=0.91\textwidth]{ 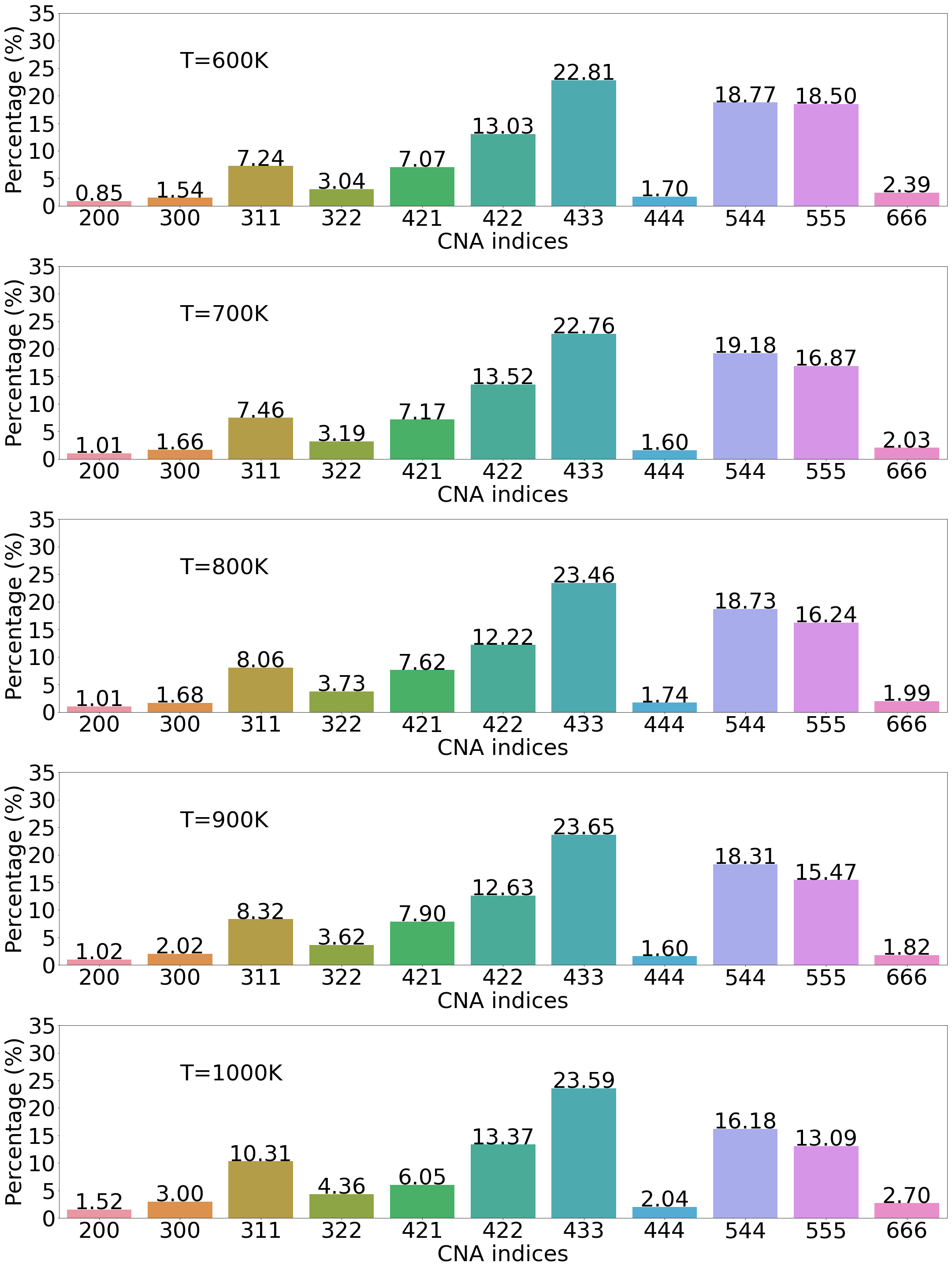} \hfill	
	\end{center}
	\caption{Abundance of pairs from the CNA centred on Al atoms in the Al$_{90}$Si$_{10}$ binary liquid alloy at different temperatures.}
	
\end{figure}

\begin{figure}[h]
	\begin{center}
		\includegraphics[angle=0 ,width=0.91\textwidth]{ 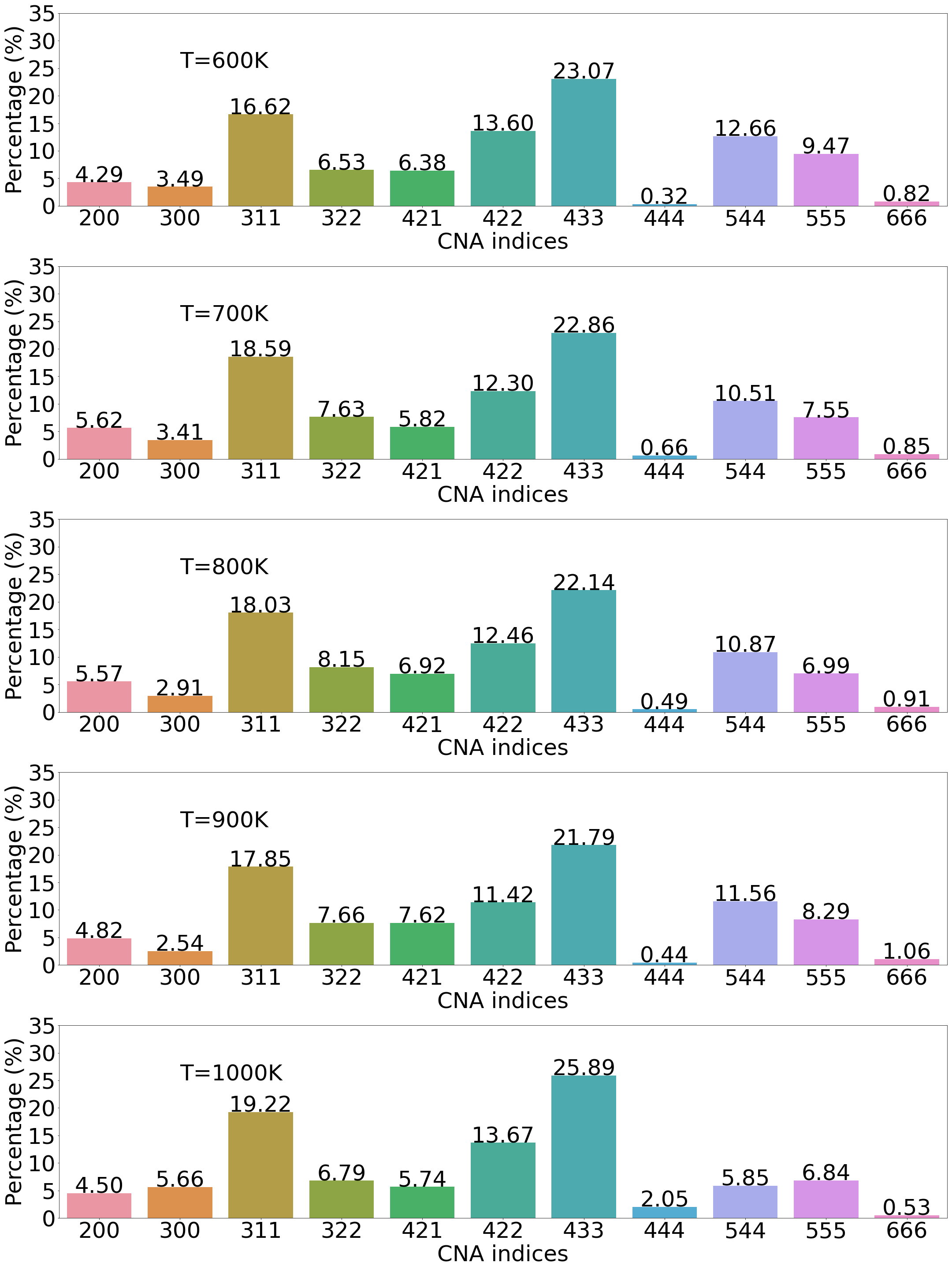} \hfill	
	\end{center}
	\caption{Abundance of pairs from the CNA centred on Si atoms in the Al$_{90}$Si$_{10}$ binary liquid alloy at different temperatures.}
	
\end{figure}	

\begin{figure}[h]
	\begin{center}
		\includegraphics[angle=0 ,width=0.91\textwidth]{ 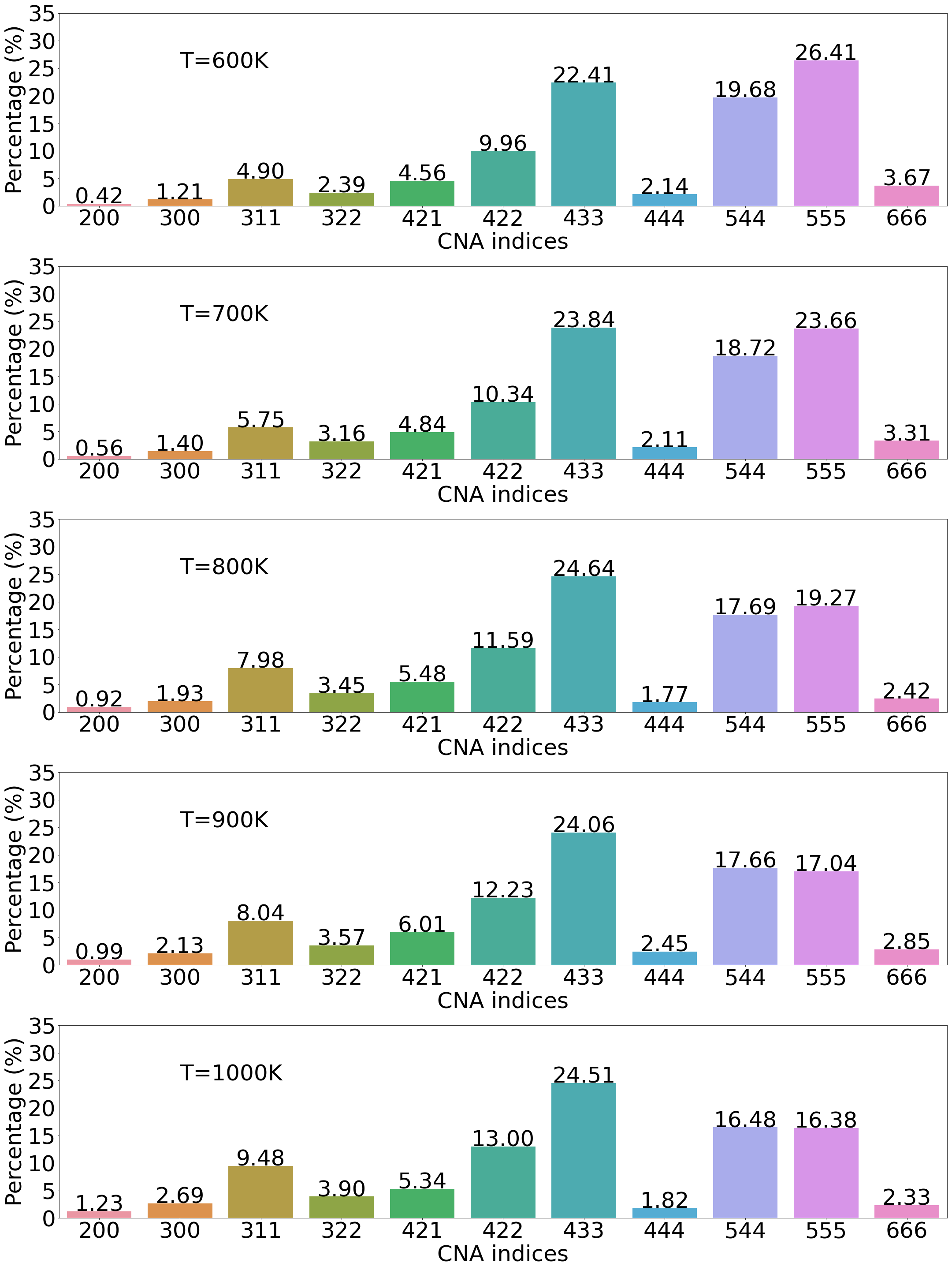} \hfill	
	\end{center}
	\caption{Abundance of pairs from the CNA centred on Al atoms in the Al$_{80}$Mg$_{10}$Si$_{10}$ ternary liquid alloy at different temperatures.}
	
\end{figure}

\begin{figure}[h]
	\begin{center}
		\includegraphics[angle=0 ,width=0.91\textwidth]{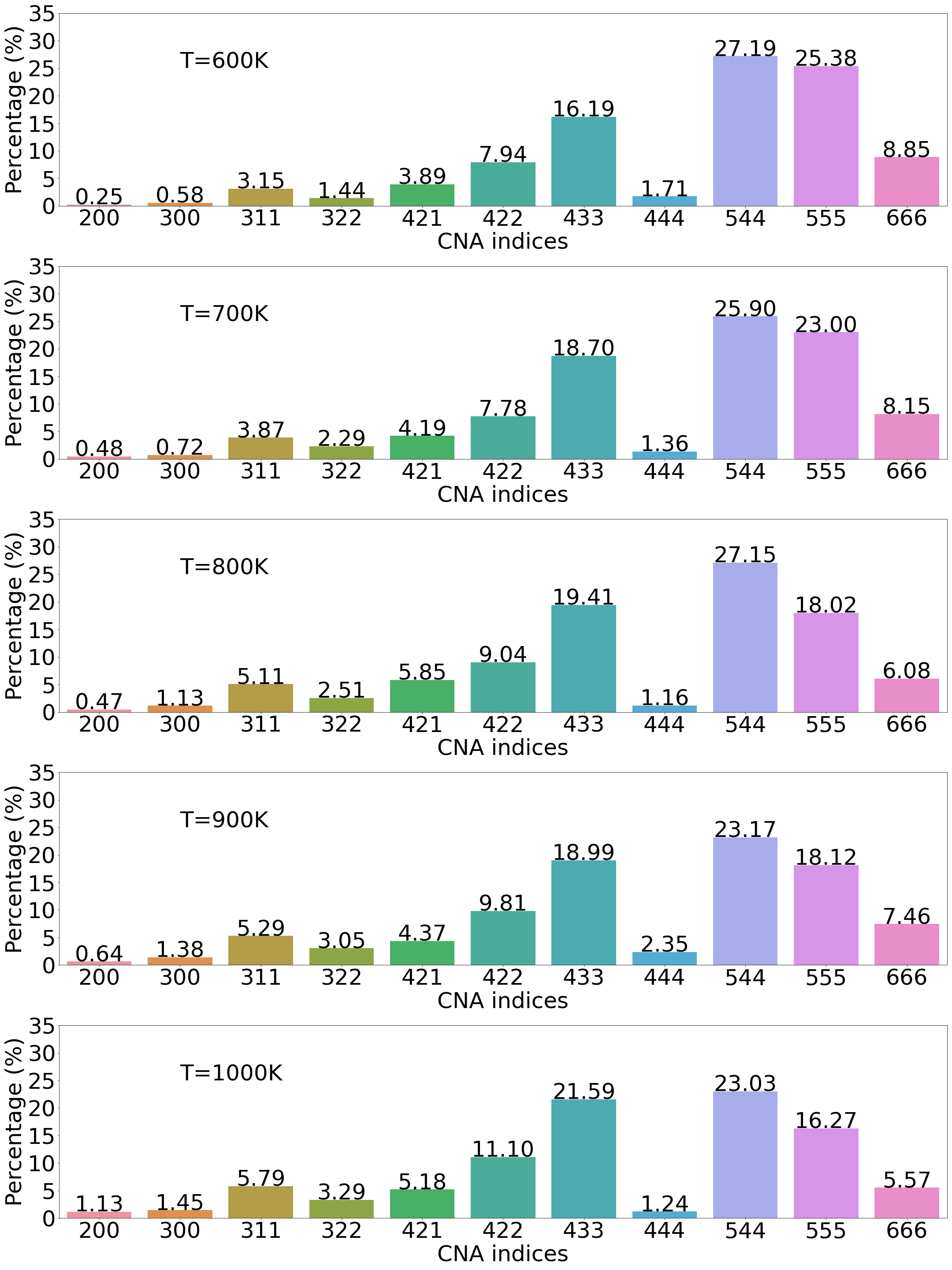} \hfill	
	\end{center}
	\caption{Abundance of pairs from the CNA centred on Mg atoms in the Al$_{80}$Mg$_{10}$Si$_{10}$ ternary liquid alloy at different temperatures.}
	
\end{figure}

\begin{figure}[h]
	\begin{center}
		\includegraphics[angle=0 ,width=0.91\textwidth]{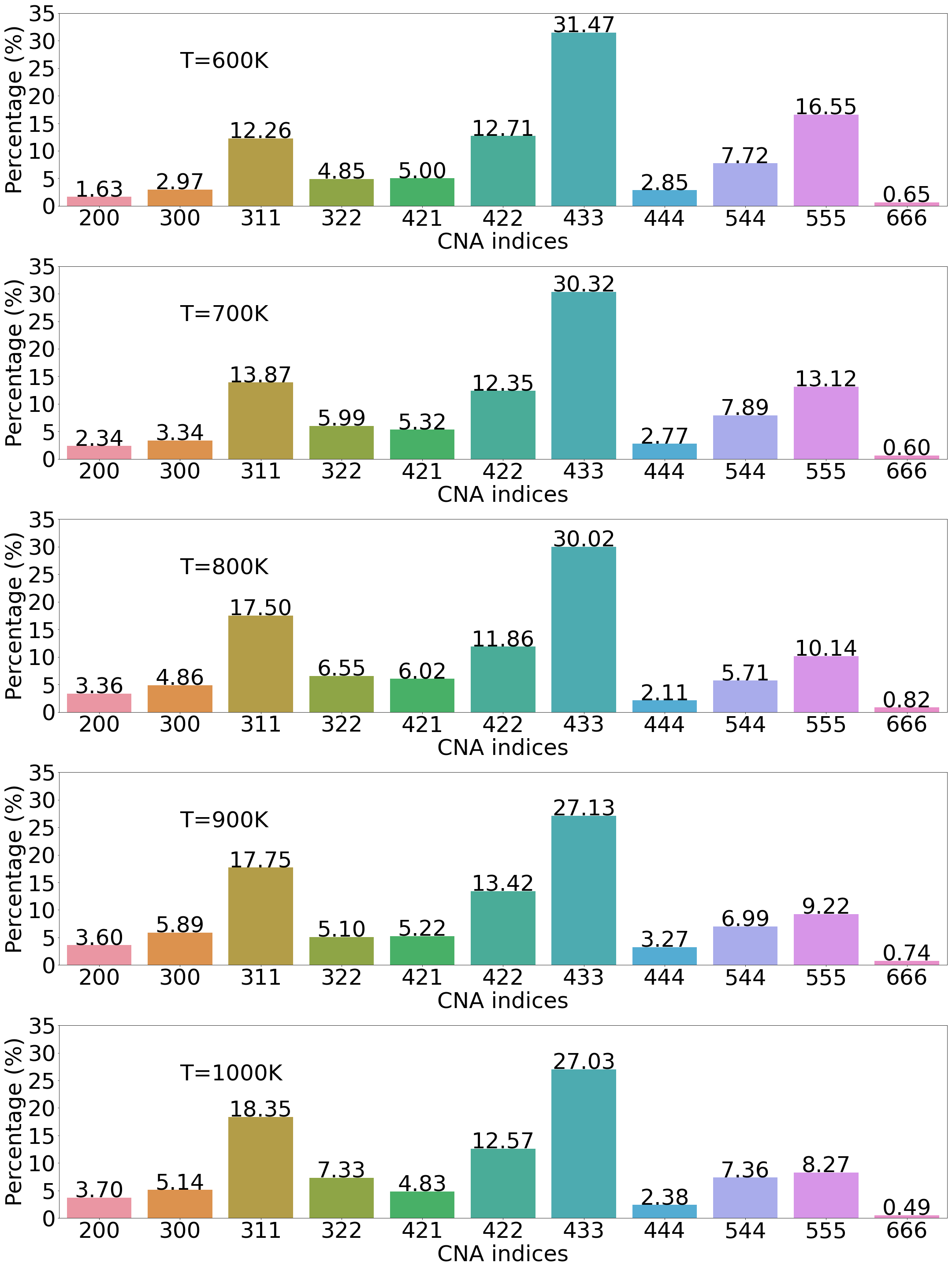} \hfill	
	\end{center}
	\caption{Abundance of pairs from the CNA centred on Si atoms in the Al$_{80}$Mg$_{10}$Si$_{10}$ ternary liquid alloy at different temperatures.}
	
\end{figure}

\begin{figure}[h]
	\begin{center}
		\includegraphics[angle=0 ,width=0.91\textwidth]{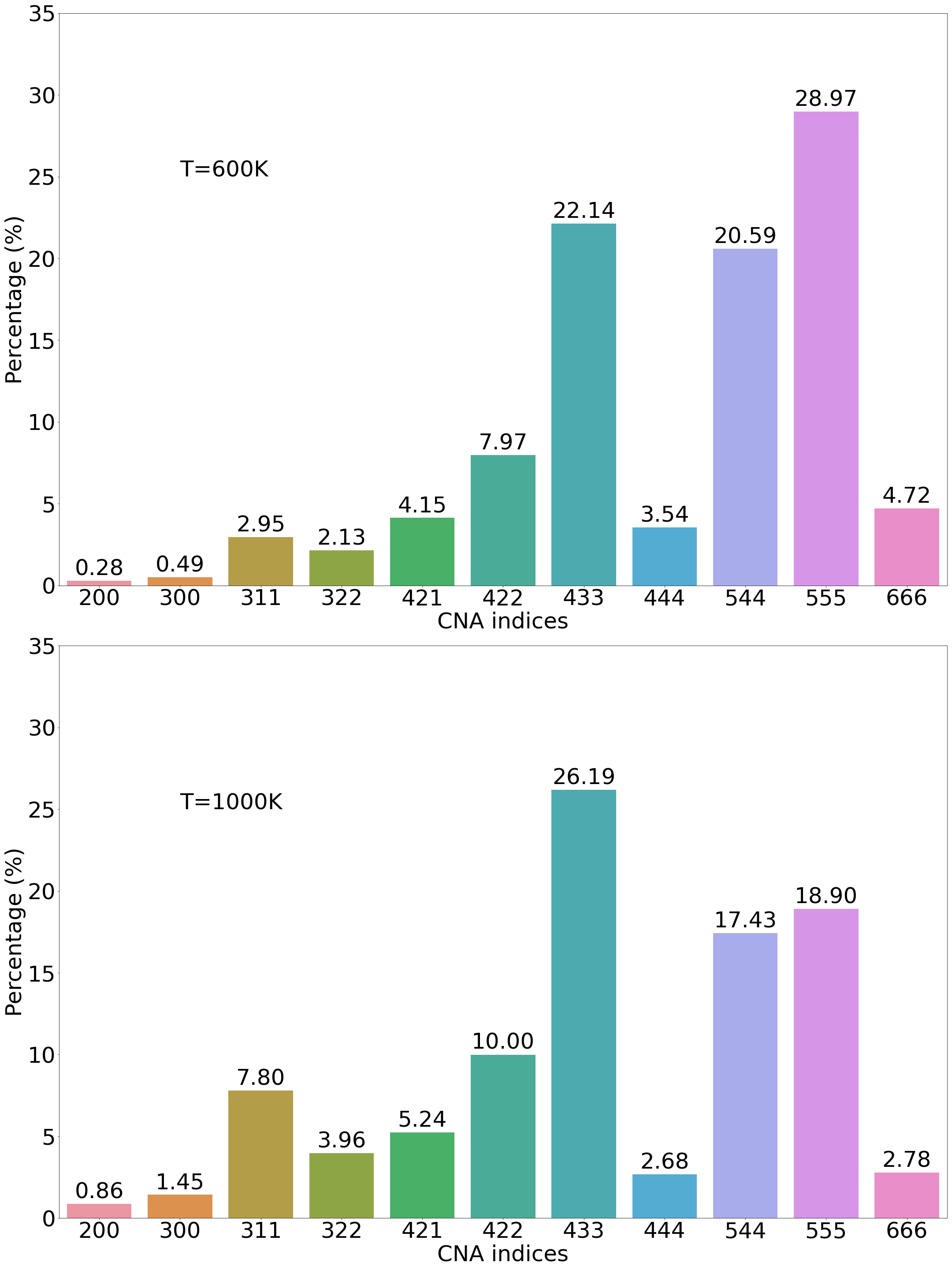} \hfill	
	\end{center}
	\caption{Abundance of pairs from the CNA centred on Al atoms in the Al$_{70}$Mg$_{20}$Si$_{10}$ ternary liquid alloy at different temperatures.}
	
\end{figure}	

\begin{figure}[h]
	\begin{center}
		\includegraphics[angle=0 ,width=0.91\textwidth]{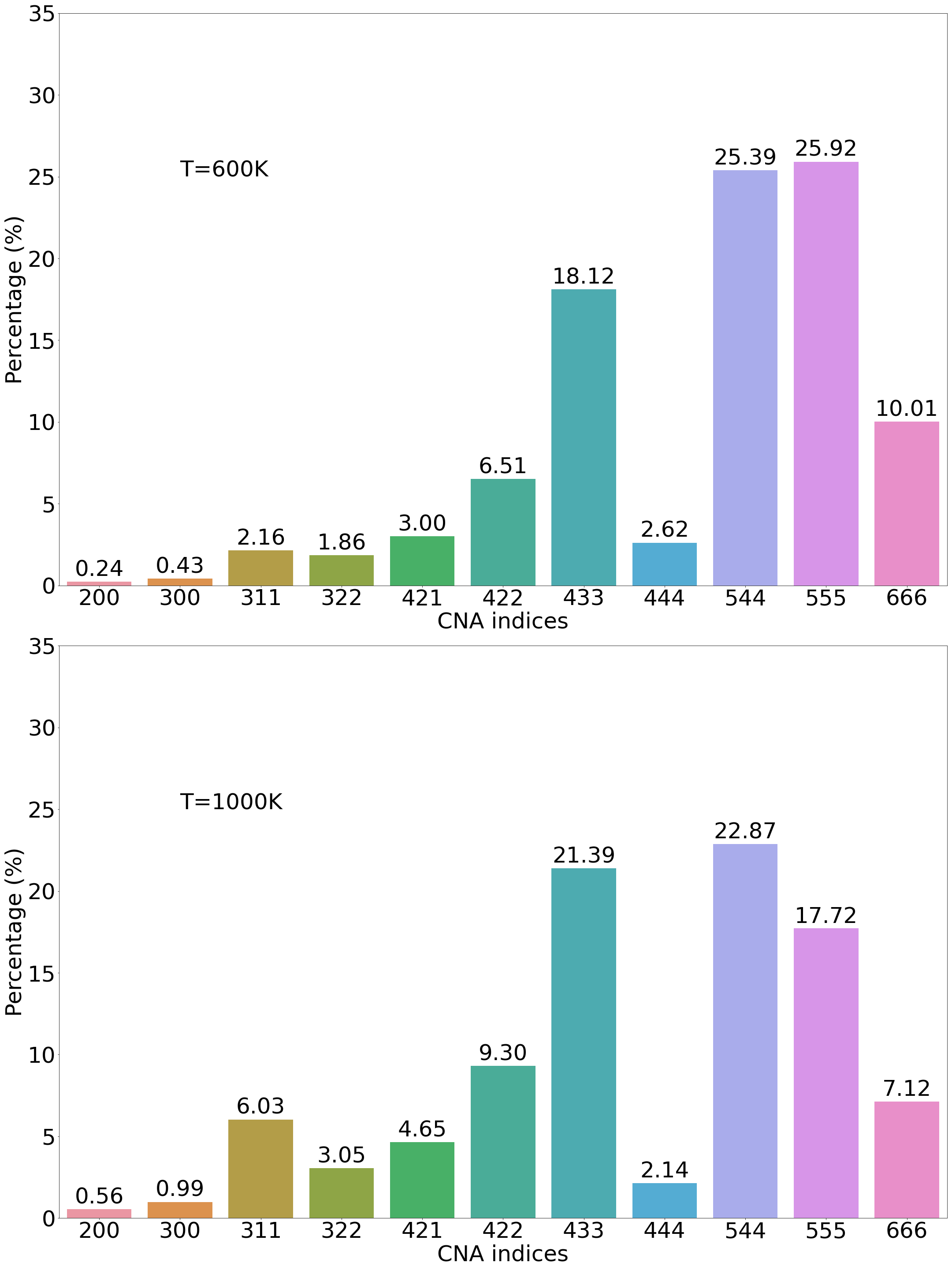} \hfill	
	\end{center}
	\caption{Abundance of pairs from the CNA centred on Mg atoms in the Al$_{70}$Mg$_{20}$Si$_{10}$ ternary liquid alloy at different temperatures.}
	
\end{figure}

\begin{figure}[h]
	\begin{center}
		\includegraphics[angle=0 ,width=0.91\textwidth]{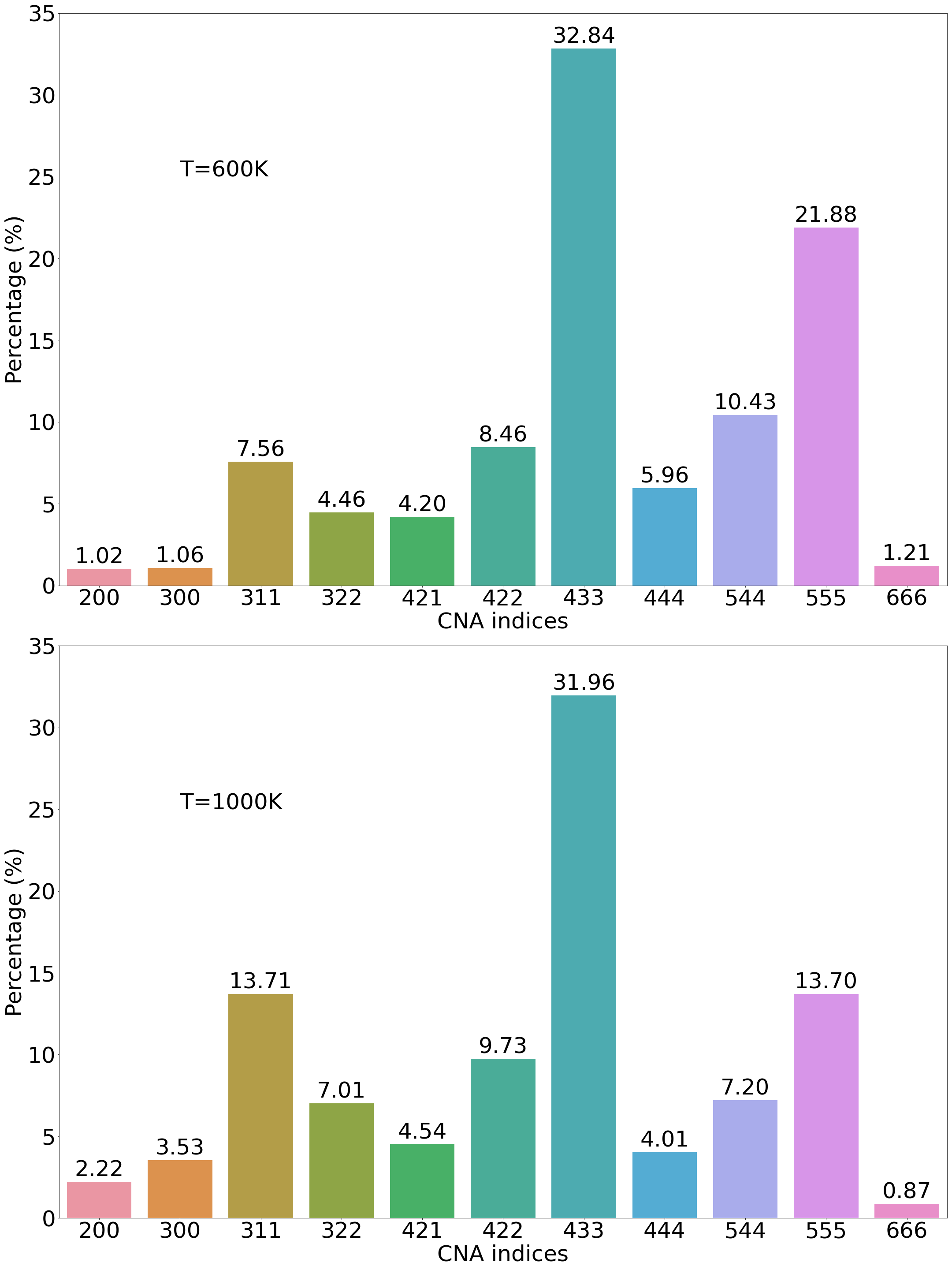} \hfill	
	\end{center}
	\caption{Abundance of pairs from the CNA centred on Si atoms in the Al$_{70}$Mg$_{20}$Si$_{10}$ ternary liquid alloy at different temperatures.}
	
\end{figure}

\begin{figure}[h]
	\begin{center}
		\includegraphics[angle=0 ,width=0.91\textwidth]{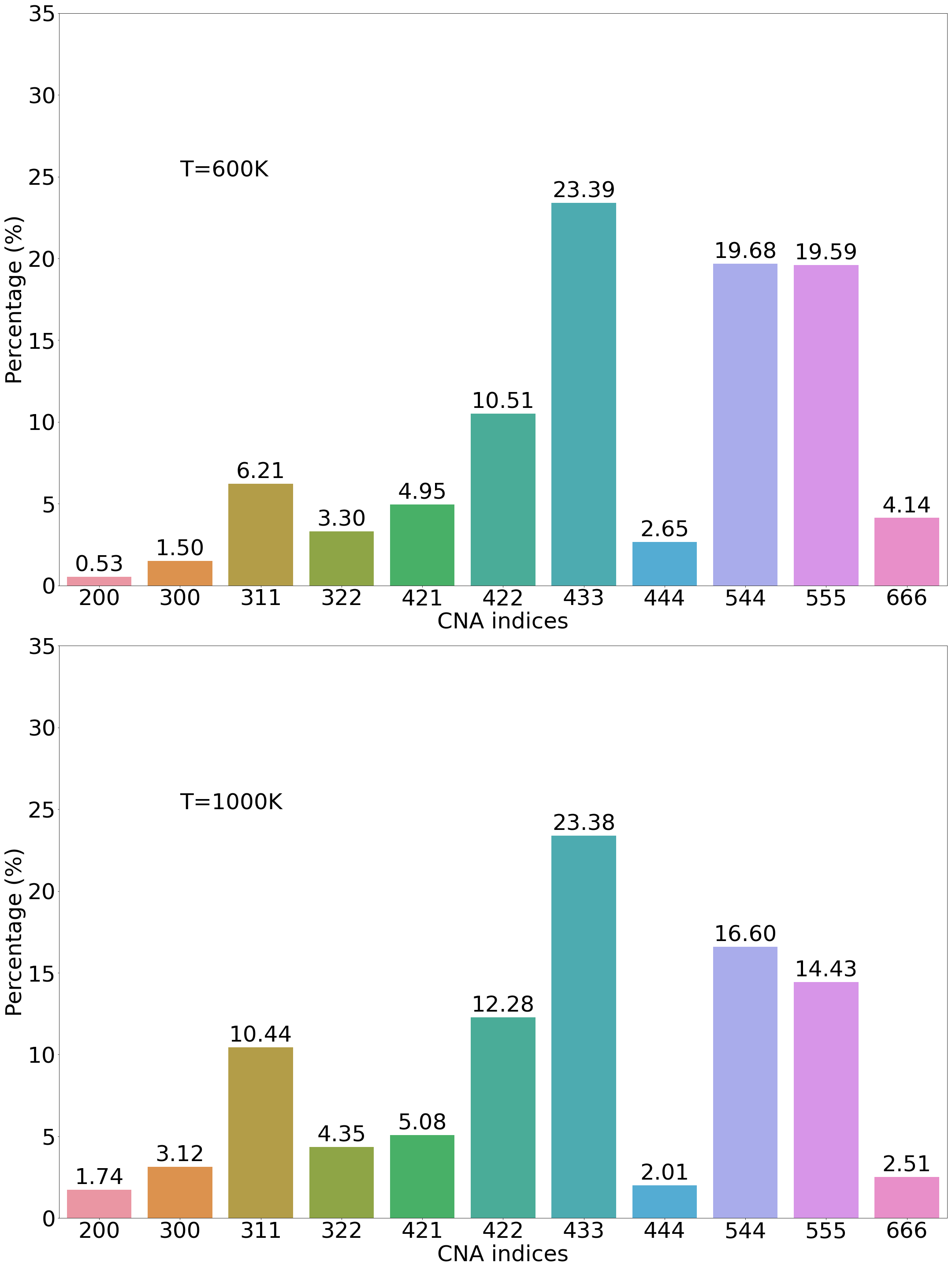} \hfill	
	\end{center}
	\caption{Abundance of pairs from the CNA centred on Al atoms in the Al$_{70}$Mg$_{10}$Si$_{20}$ ternary liquid alloy at different temperatures.}
	
\end{figure}	

\begin{figure}[h]
	\begin{center}
		\includegraphics[angle=0 ,width=0.91\textwidth]{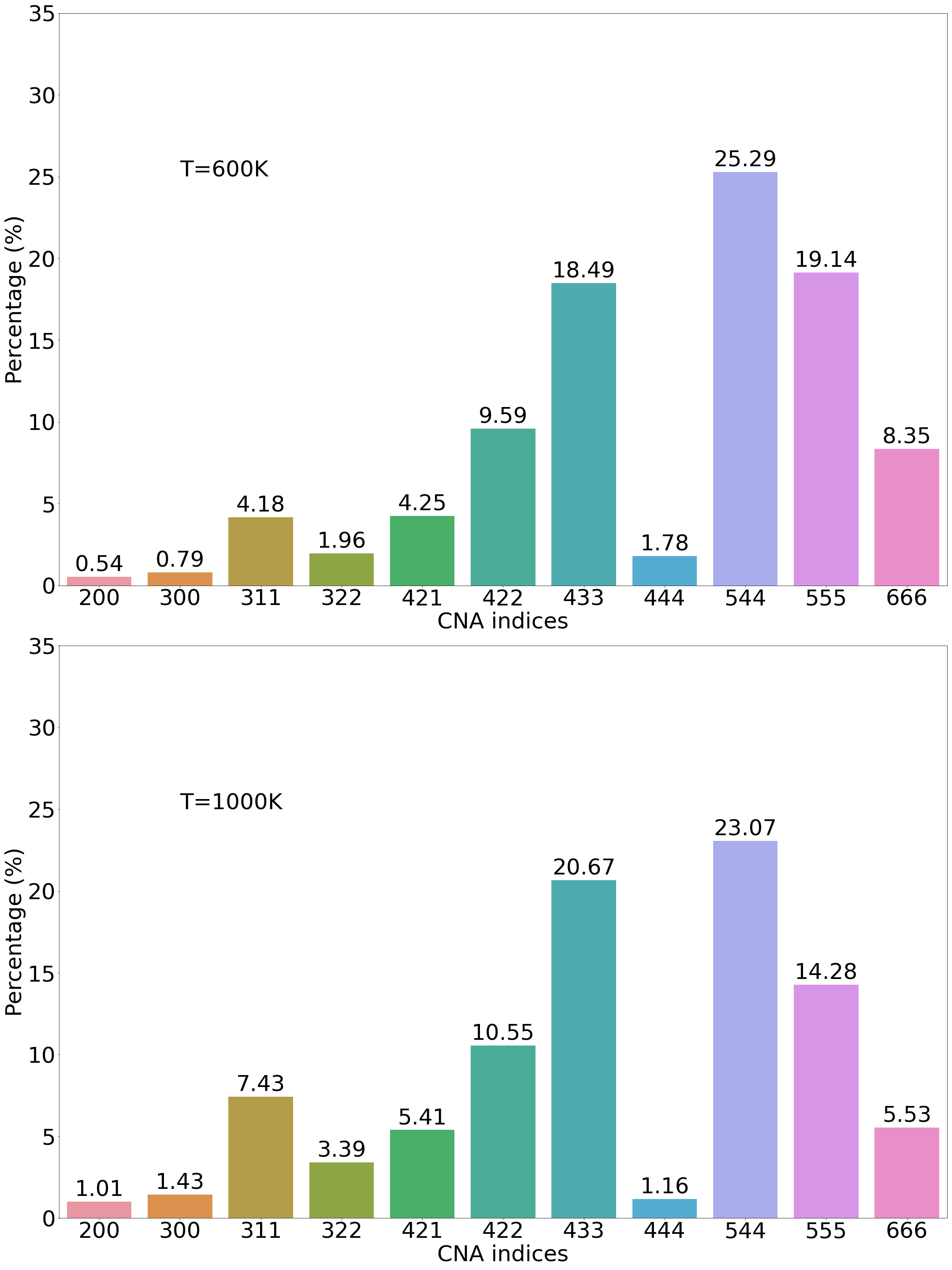} \hfill	
	\end{center}
	\caption{Abundance of pairs from the CNA centred on Mg atoms in the Al$_{70}$Mg$_{10}$Si$_{20}$ ternary liquid alloy at different temperatures.}
	
\end{figure}

\begin{figure}[h]
	\begin{center}
		\includegraphics[angle=0 ,width=0.91\textwidth]{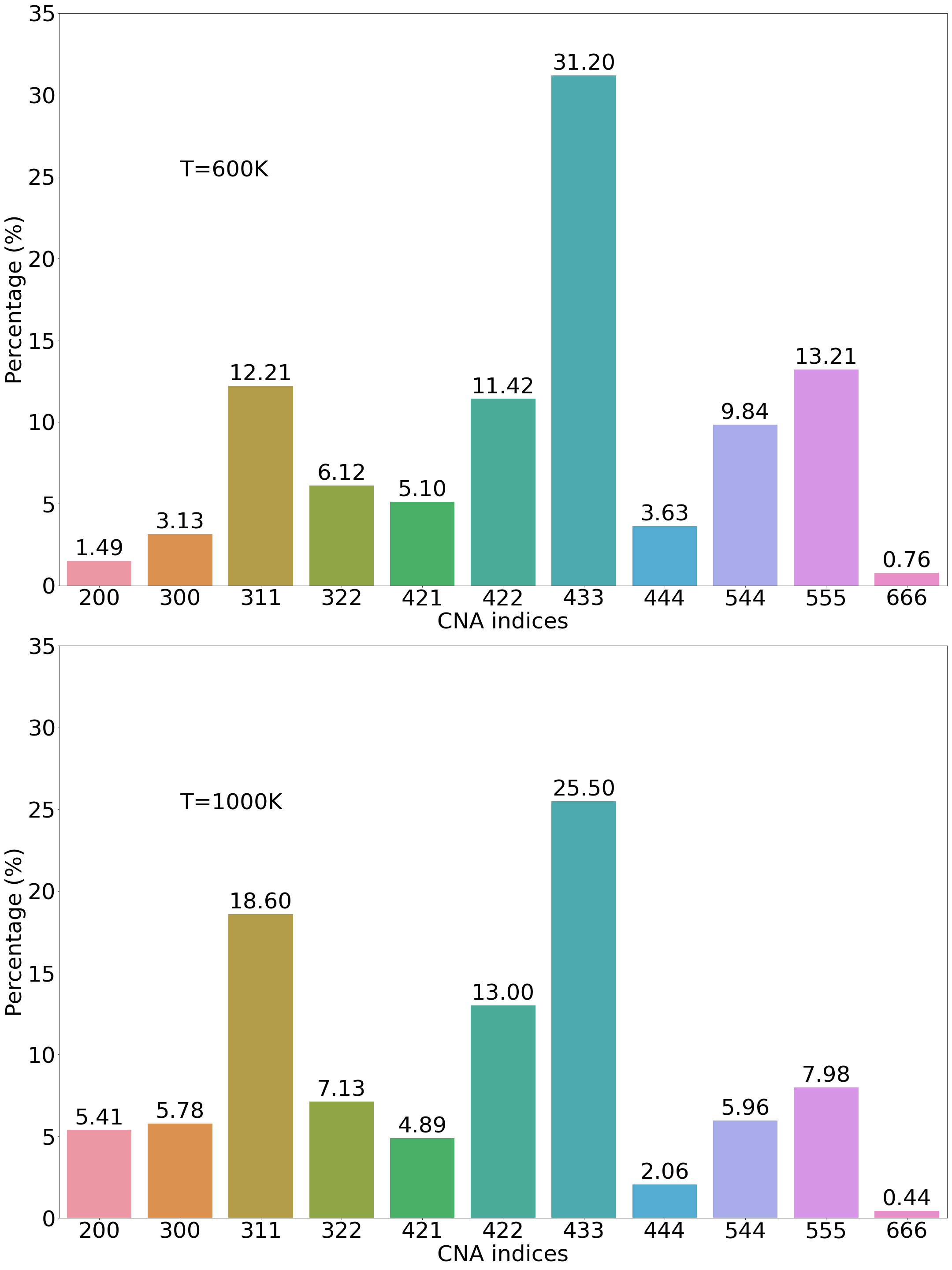} \hfill	
	\end{center}
	\caption{Abundance of pairs from the CNA centred on Si atoms in the Al$_{70}$Mg$_{10}$Si$_{20}$ ternary liquid alloy at different temperatures.}
	
\end{figure}

\begin{figure}[h]
	\begin{center}
		\includegraphics[angle=0 ,width=1\textwidth]{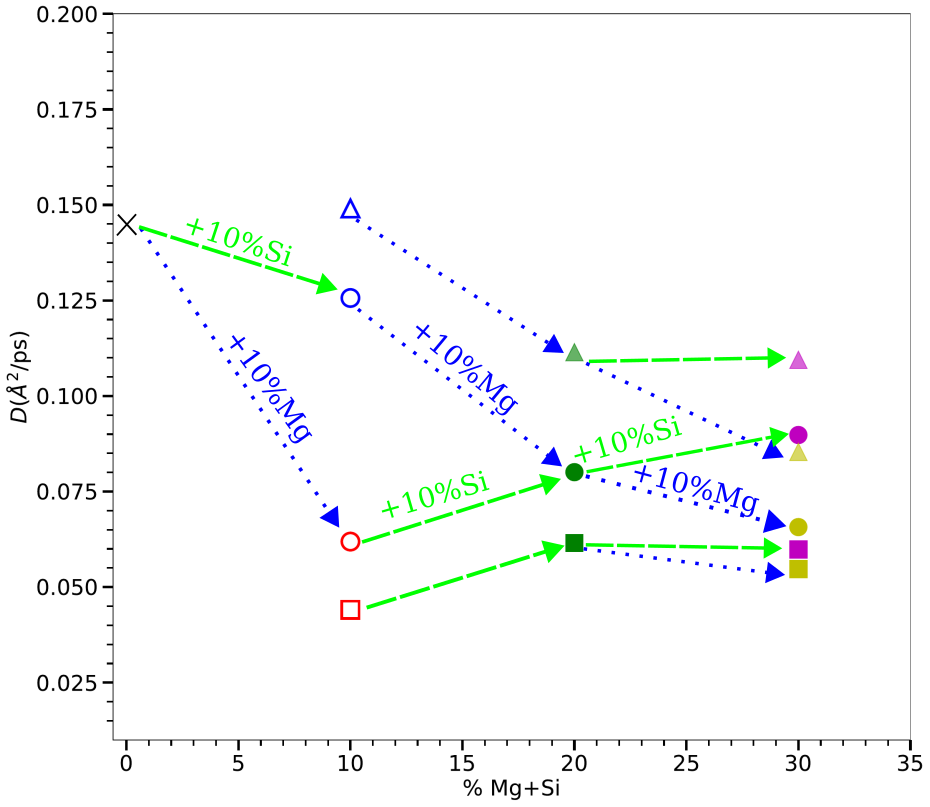} 
	\end{center}
	\caption{Diffusion coefficients for all the liquid alloys as function of composition at $T=$600 K. The open symbols represent the binary liquid alloys (Red: Al$_{90}$Mg$_{10}$; Blue: Al$_{90}$Si$_{10}$) and the closed symbols represent ternary liquid alloys (Green: Al$_{80}$Mg$_{10}$Si$_{10}$; Dark yellow: Al$_{70}$Mg$_{20}$Si$_{10}$; Magenta: Al$_{70}$Mg$_{10}$Si$_{20}$). The cross represent the value for pure liquid Al from \cite{Jakse2023}.
	}
	
\end{figure}

%
%
%
%
%
%

\end{document}